\documentclass[preprint,12pt]{elsarticle}

\usepackage{amssymb}
\usepackage{amsmath}
\usepackage{longtable}
\usepackage{bbm}

\usepackage{booktabs}
\usepackage{siunitx}
\usepackage{makecell}
\usepackage{lineno}
\usepackage{algorithm}
\usepackage{algorithmic}
\usepackage{subcaption}
\usepackage{tabularx} 

\journal{Knowledge-Based Systems}

\begin{document}

\begin{frontmatter}



\title{Botfip-LLM: An Enhanced Multimodal Scientific Computing Framework Leveraging Knowledge Distillation from Large Language Models}

 \author[inst1]{Tianhao Chen}
 \author[inst2,inst3]{Pengbo Xu}
 \author[inst1]{Haibiao Zheng}
 
 \affiliation[inst1]{organization={School of Mathematical Sciences, Key Laboratory of MEA (Ministry of Education), Shanghai Key Laboratory of PMMP, East China Normal University}, 
 	addressline={Shanghai 200241}, 
 	city={Shanghai},
 	postcode={200241}, 
 	country={P.R. China}}
 
 \affiliation[inst3]{organization={School of Mathematical Sciences, Key Laboratory of MEA (Ministry of Education), Shanghai Key Laboratory of PMMP, East China Normal University}, 
 	addressline={Shanghai 200241}, 
 	city={Shanghai},
 	postcode={200241}, 
 	country={P.R. China}}
 
 \affiliation[inst4]{organization={Shanghai Zhangjiang Institute of Mathematics}, 
 	addressline={Shanghai}, 
 	city={Shanghai},
 	postcode={201203}, 
 	country={China}}
\begin{abstract}

In recent years, the introduction of AI technologies has brought transformative changes to scientific computing. However, AI models typically focus on single-task and single-modal data processing, limiting their application. To address this, multimodal scientific computing frameworks have become a trend. The Botfip framework aligns function images with symbolic operation trees through multimodal training, extracting deep scientific information. However, Botfip struggles with processing Formula Strings, leading to inadequate understanding in multimodal learning. To enhance Botfip's learning of Formula Strings and expand its applicability to related tasks, we propose the Botfip-LLM framework based on knowledge distillation, incorporating pre-trained large language models for aligning symbolic tree data. Experimental analysis shows that the choice of LLM is crucial, with ChatGLM-2 outperforming others in training and testing. Botfip-LLM not only improves performance, generalization, and extrapolation over the original Botfip model but also significantly enhances applicability to Formula String-related tasks, enabling more diverse task handling.
\end{abstract}

\begin{graphicalabstract}
\centering
\includegraphics[width=0.82\textwidth]{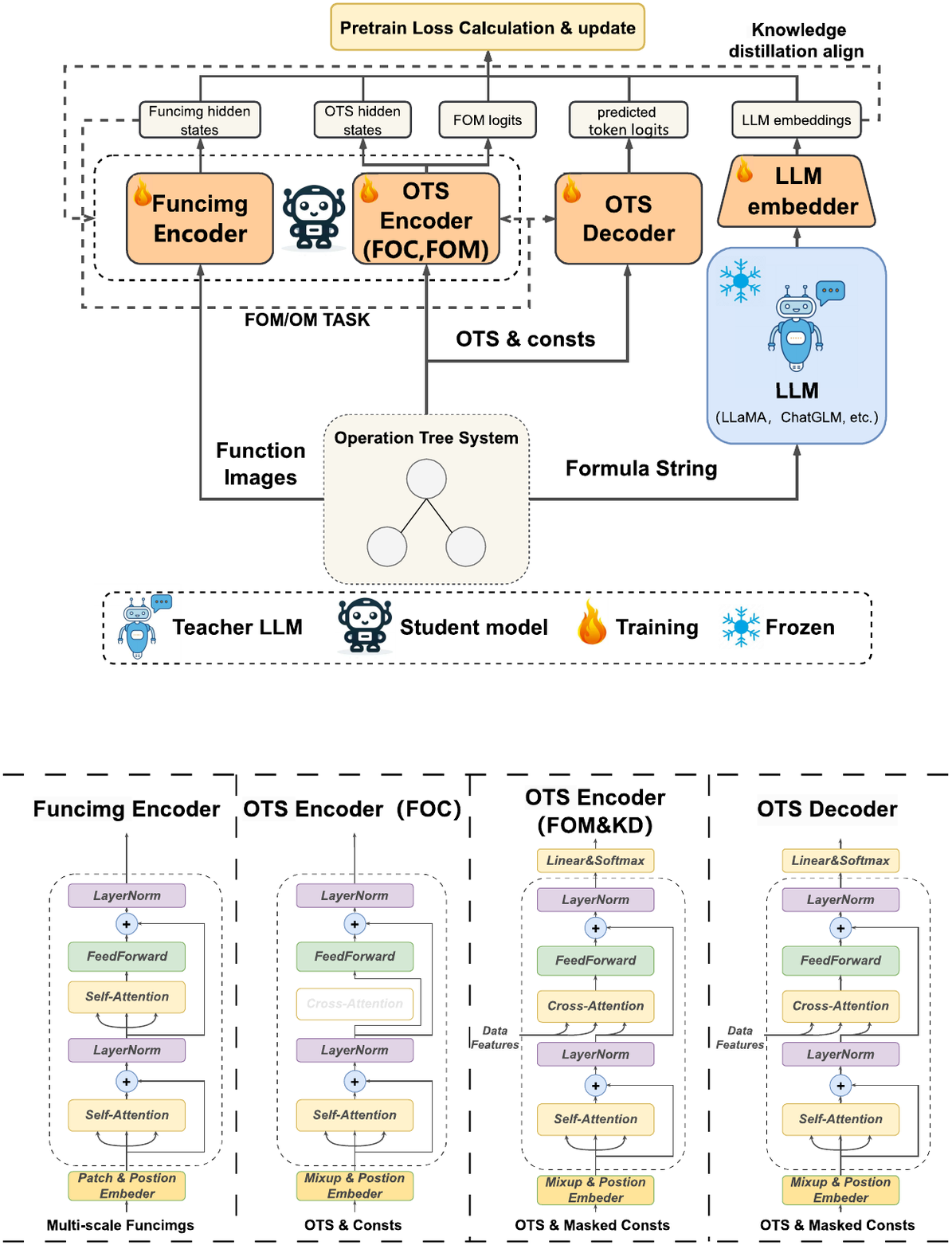}
\end{graphicalabstract}

\begin{highlights}
\item \textbf{An Innovative Multimodal Scientific Computing Framework}: Botfip-LLM is an innovative multimodal scientific computing framework that integrates numerical information (function images), sequence encoding information, and symbolic information (symbolic formulas) centered around symbolic operation trees. This integration not only captures multiple dimensions of scientific data but also enhances the model's capability to understand and handle complex scientific phenomena through the synergistic effect of multimodal information.

\item \textbf{Knowledge Distillation in Botfip-LLM}: Botfip-LLM employs knowledge distillation to infuse the experience and knowledge of pre-trained large language models (LLMs) into the scientific computing multimodal framework. This process not only improves the model's understanding of multimodal information but also significantly enhances its performance in scientific computing tasks by efficiently transferring the deep features of the pre-trained models, thereby boosting the fusion of multimodal data.

\item \textbf{Efficient Training with Distributed Deployment and Aggregation}: Botfip-LLM successfully achieves the invocation of LLMs and the training of the main model in low GPU memory environments through distributed deployment and aggregation. This approach addresses the challenge of deploying large models in resource-constrained settings while ensuring the efficiency and scalability of the training process, making the application of large-scale models more practical and feasible.

\item \textbf{Expansion of Downstream Tasks and Potential Integration with NLP}: Building on the Botfip framework, Botfip-LLM can extend to more downstream tasks, such as generating and inferring symbolic sequence encodings from symbolic formulas. In the future, this framework can further integrate with various natural language processing (NLP) tasks, exploring more cross-domain application scenarios and innovative research directions, thus promoting the convergence of scientific computing and NLP.
\end{highlights}


\end{frontmatter}



\section{Introduction}
\label{sec:Introducion}

In recent years, the exponential growth in data complexity and volume has posed significant challenges to traditional scientific methods. As computational challenges become increasingly complex and data-intensive, traditional tools, though powerful, often struggle to keep pace with the scale and complexity of contemporary scientific datasets. This situation urgently calls for more robust, efficient, and scalable solutions, paving the way for the rise of artificial intelligence (AI) in scientific computing. Integrating AI technologies into scientific computing marks a transformative shift in research and development across multiple scientific fields, such as weather forecasting \cite{conti2024artificial,valanarasu2022transweather,ji2024spatio}, numerical solutions for partial differential equations (PDEs) \cite{cai2021physics,raissi2019physics,li2020fourier,hao2023gnot}, genomics \cite{khan2021gene,zhang2022transformer}, drug analysis \cite{kolluri2022machine,chen2023artificial,fu2022spectratr}, and symbolic regression \cite{valipour2021symbolicgpt,vastl2024symformer,kamienny2022end}. The fusion of AI with scientific inquiry not only enhances existing computational methods but also opens new avenues for exploration and discovery, giving rise to the rapidly evolving field known as “AI for Science” \cite{stevens2020ai}, which is dedicated to leveraging AI technologies to tackle complex challenges in scientific research.

Despite the progress that AI has made in scientific research, its potential applications remain constrained by several pervasive issues. The primary challenge is that AI models typically focus on single-task and single-modal data processing, which limits their ability to fully grasp complex scientific phenomena, especially in studies involving multifactor interactions. These single-task-oriented models also struggle to generalize across different scientific problems, hindering the development of robust and versatile models. Furthermore, the prevalent use of single-modal data processing overlooks the richness and interconnectedness of scientific data. To address these issues, the development of multimodal scientific computing frameworks has become an inevitable trend. As an emerging field, multimodal learning aims to integrate various data forms, such as text, images, and sound, to achieve complementarity between data types and enhance the overall performance and applicability of AI models. This integrative approach aids in understanding complex phenomena and promotes the application of models in multitask scenarios. For instance, OpenAI's Contrastive Language-Image Pretraining Model (CLIP) \cite{radford2021learning} combines text and image understanding to provide linguistic support for visual tasks. Additionally, multimodal learning has achieved significant accomplishments in computer vision and text processing, and is now expanding into fields such as scientific computing. Other key developments in this field include BoOTStrapping Language-Image Pre-training Model (BLIP) \cite{li2022blip}, ALIGN \cite{jia2021scaling}, and ALBEF \cite{li2021align}, among others. However, despite the notable advances in multimodal learning within computer vision and text processing, its application in the AI for Science domain remains in its nascent stages. Nonetheless, there have been some representative works, such as in \cite{chen2024boOTStrapping}, where researchers proposed the Botfip multimodal scientific computing framework based on the Blip model. Botfip integrates multi-scale function images and Operation Tree Skeleton Sequence (OTS) features to explore the deep connections between function images and their corresponding symbolic expression sequences, applying these insights to downstream tasks like symbolic regression. This innovative approach not only marks a new application of multimodal learning in scientific computing but also demonstrates its potential transformative nature in handling complex scientific data.

\begin{figure}[!t]
    \centering
    \includegraphics[width=1.0\textwidth]{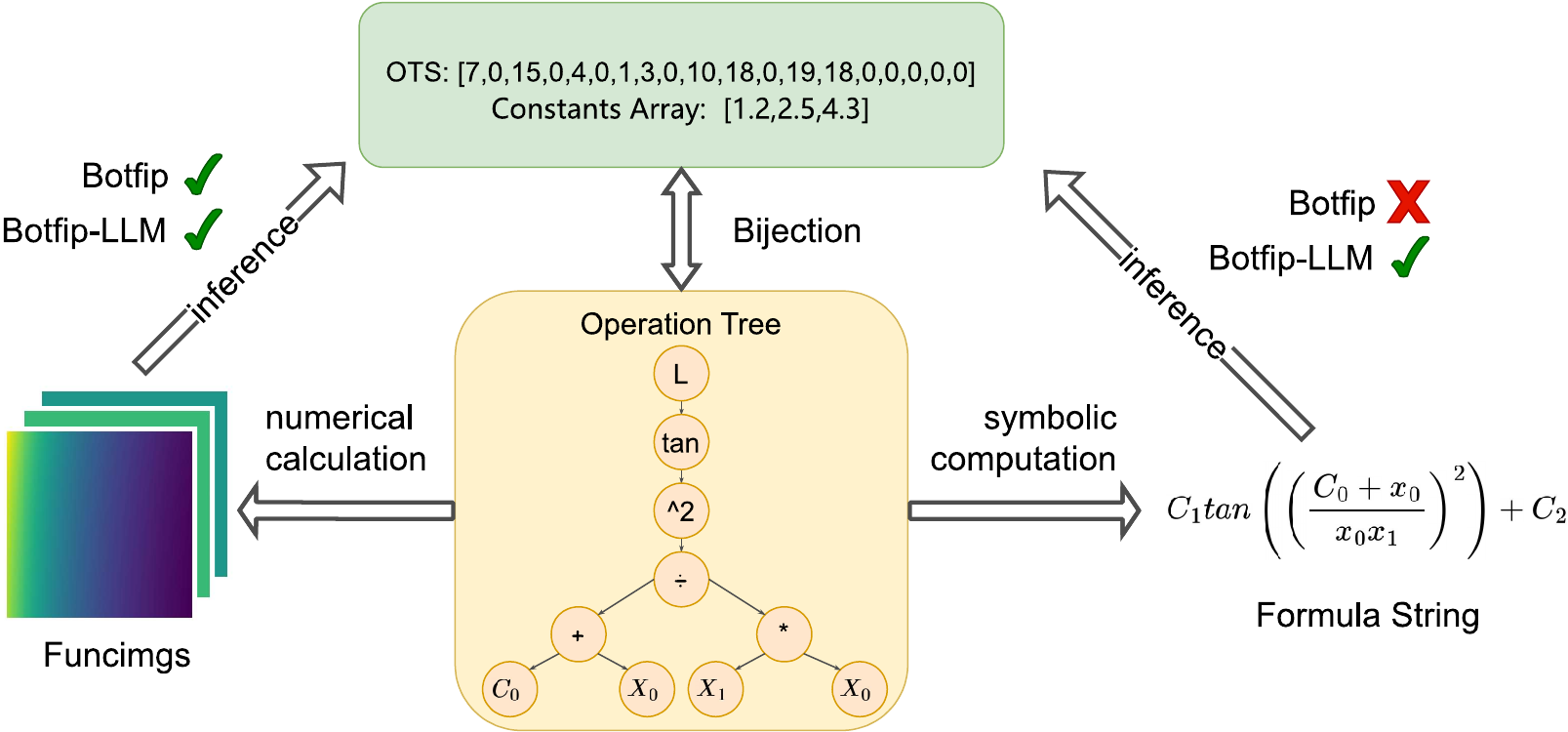}
    \caption{Multimodal data representation of Funcimg-OTS-Formula String in scientific computing. While Funcimg and Formula String cannot be directly converted to the corresponding OTS of the symbolic operation tree, the Botfip framework facilitates the conversion of Funcimg to OTS through multimodal training but cannot directly convert Formula String.}
    \label{fig:botfip_limit}
\end{figure}

Despite Botfip's use of a Multi-Encoder-Decoder (MED) architecture, which significantly enhances its generalization, versatility, and applicability compared to single-modal, single-task scientific computing models, its reliance on sequence encoding networks such as the BERT model \cite{devlin2018bert} and image encoding networks like Vision Transformer (ViT) \cite{dosovitskiy2020image} limits its parameter scale and thus its extrapolation capability. Currently, large language models (LLMs) such as LLaMA \cite{touvron2023llama} and Gemma \cite{team2024gemma}, due to their vast parameter scales and deep network structures, are more effective at handling and generating complex data patterns. These large models have demonstrated substantial advantages over smaller models like BERT in fields such as natural language processing (NLPs), becoming the mainstream focus of research. Additionally, multimodal text-image models have started leveraging large pre-trained models to enhance the joint representation of images and texts, improving the accuracy of text-image matching and generation techniques. For example, the BLIP-2 \cite{li2023blip} model combines well-trained vision models and LLMs. With the powerful generalization ability, knowledge base, and comprehension and generation capabilities of LLMs, BLIP-2 achieves leading results through a lightweight query transformer, showcasing the feasibility and potential of integrating LLMs into multimodal learning.

\begin{figure}[!t]
    \centering
        \subfloat[Visualization of the Botfip-LLM Framework]{
        \includegraphics[width=0.8\textwidth]{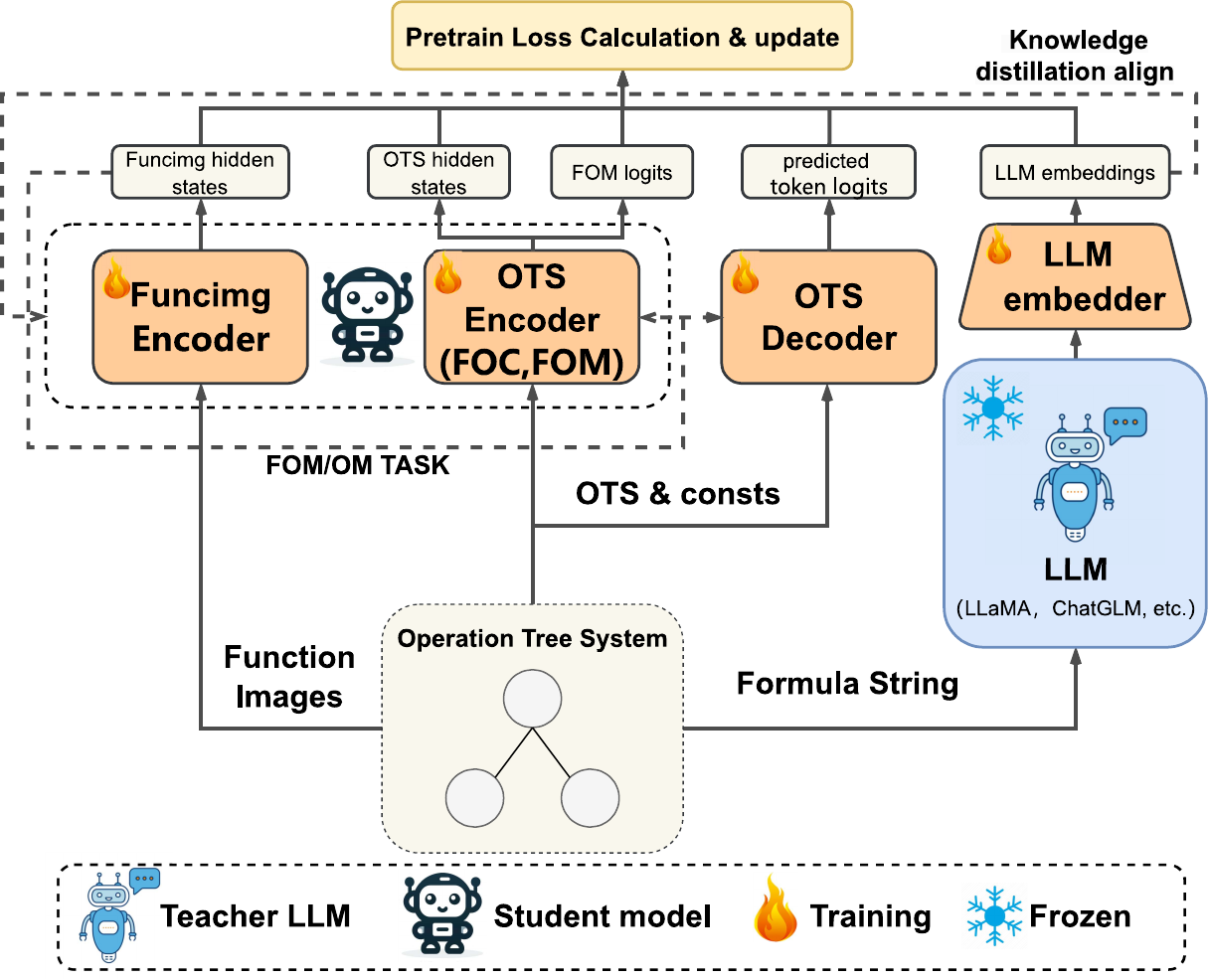}
        \label{fig:Botfip_LLM_framework}
    }\\ %
            \subfloat[Visualization of Block Details in the Botfip-LLM Framework]{
        \includegraphics[width=0.8\textwidth]{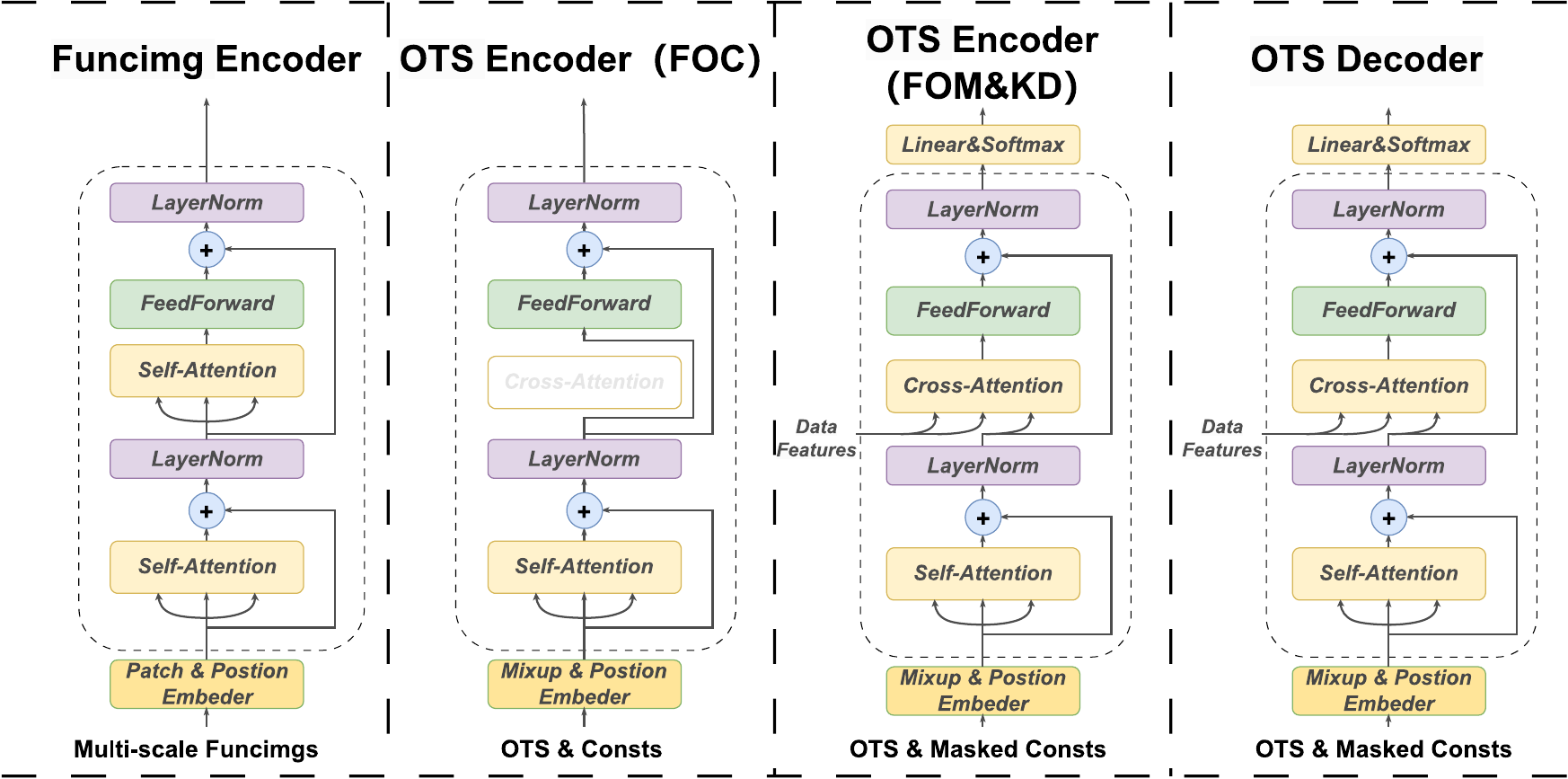}
        \label{fig:Block structure}
    }
    \caption{Visualization of the main structure and related details of the Botfip-LLM framework, with Figure \ref{fig:Botfip_LLM_framework} illustrating the main forward computation and operational process during the pre-training phase, and Figure \ref{fig:Block structure} providing detailed information and forward computation processes of each major module.}
    \label{fig:Botfip_total_LLM_framework}
\end{figure}

However, LLMs are currently not suitable for direct replacement of foundational architectures like BERT in multimodal scientific computing frameworks such as Botfip. First, LLM training requires massive data support. Unlike fields like NLPs, scientific computing has far fewer general datasets available. Although \cite{chen2024boOTStrapping} proposed a method for random symbolic operation tree generation and a general scientific computing data format (function image-OTS pairs) to generate large amounts of data for pre-training, the limited audience and data sources in scientific computing cannot match the scale of the text-image domain. Additionally, the original Botfip framework primarily involves multi-scale function images and OTS, without addressing the symbolic expressions corresponding to the function-related symbolic operation trees for information processing and feature extraction. This is because OTS and constant function vectors can directly convert into computable symbolic operation trees, reconstructing the corresponding symbolic formulas. However, this approach limits the thorough exploration of symbolic formula information. Using only OTS to represent symbolic information is flawed because different OTS and constant vector pairs can correspond to the same symbolic formula (e.g., swapping two commutative operations with parameters in the symbolic operation tree), and solely relying on OTS as the symbolic information source also hinders the introduction of LLMs. Thus, the Botfip framework cannot, like multimodal models such as BLIP-2, directly utilize pre-trained LLMs and existing extensive pre-trained datasets to reduce training costs and quickly adapt and optimize for various tasks, thereby eliminating the need to train models from scratch. This also means that forcibly replacing small parameter models like BERT with large models would significantly increase Botfip's training costs, requiring substantial time and computational resources for some tasks, presenting a huge challenge. How to effectively apply large model methods to Botfip and similar multimodal scientific computing frameworks, and the broader AI for Science domain, is a pressing issue that needs to be addressed.

\textbf{Contribution:} In this paper, to overcome the limitations imposed by the complexity and computational cost of directly applying LLMs in the Botfip framework, address the deficiency of Botfip in fully utilizing symbolic formula information, and further extend the functionalities of the Botfip model, we propose the Botfip-LLM enhanced extension framework based on knowledge distillation technology. The Botfip-LLM enhanced extension framework builds on Botfip by additionally employing pre-trained LLMs to handle the symbolic expressions generated from the OTS corresponding to the symbolic operation tree. These expressions are transformed into high-dimensional hidden states (or logits) by the LLMs. After freezing the parameters of the LLMs, they do not participate in the training of the Botfip model. The resulting encoded vectors are dimensionally reduced through the LLMs' corresponding embedder and then trained via contrastive learning with the encoded vectors obtained from the function image encoder and OTS encoder. This achieves effective alignment of the high-dimensional semantic representations generated by the LLMs with the multimodal input data in the Botfip model. This process can be referred to as knowledge distillation guided by LLMs.

This alignment process involves not only the standardized processing of text and image data but also the feature vectorization of symbolic formulas, ensuring that data from different modalities can interact and learn within the same semantic space. Pre-trained LLMs are typically trained on extensive text and data sets, enabling them to capture rich semantic features and complex data patterns. By transmitting these advanced feature extraction capabilities to the Botfip encoders through the alignment process, the Botfip model's feature extraction and understanding capabilities when handling specific scientific data can be significantly improved. Moreover, the pre-training process of LLMs includes handling and parsing vast amounts of data from diverse sources, endowing them with excellent generalization capabilities. Transferring this capability to the Botfip encoders through knowledge distillation can enhance Botfip's adaptability and prediction accuracy when faced with unknown or rare data.

After integrating LLMs into the Botfip framework via knowledge distillation, the downstream task range of the Botfip framework can also be further extended. For instance, the previous Botfip model primarily generated the corresponding computable symbolic operation tree directly from OTS and constant vectors, further producing function expressions. Although this process effectively reconstructs mathematical formulas from given OTS and numerical data, it is essentially unidirectional, only allowing conversion from structured OTS to function expressions but not the reverse, from function expressions to structured OTS or corresponding numerical vectors. The introduction of pre-trained large language models (LLMs) significantly improves this limitation. With the deep learning capabilities of LLMs, the Botfip-LLM framework can now not only generate symbolic expressions from OTS but also derive OTS sequences and constant vectors from function symbolic expressions. When the symbolic expressions of functions are input into the LLMs, the model first converts these expressions into high-dimensional hidden states, which encapsulate the deep semantic structure of the expressions. Subsequently, these high-dimensional states are dimensionally reduced through a specific embedder, and the resulting encoded vectors are used to guide the generation of corresponding OTS sequences and constant vectors via a CrossAttention mechanism, thereby achieving direct conversion from symbolic expressions to corresponding computable symbolic operation trees. Consequently, with the help of LLMs, Botfip-LLM can achieve mutual conversion between symbolic expressions, OTS, and function images, greatly enhancing its technical performance and applicability, and further expanding the diversity of its available downstream tasks.

\section{Related Work}
\subsection{Multi-modal Learning}

In recent years, with the continuous development of AI algorithms and computational power, multimodal learning has made significant progress in the field of artificial intelligence. As a milestone in multimodal learning, the CLIP model \cite{radford2021learning} has demonstrated strong generalization across tasks by learning from image-text pairs. It uses a simple pre-training task to learn image representations on large datasets, followed by zero-shot transfer through natural language. CLIP excels in various computer vision tasks, competing with fully supervised baselines without specific dataset training. The ALBEF model \cite{li2021align} aligns image and text representations by introducing contrastive loss and then fuses them through cross-modal attention, enhancing the learning of visual and language representations. Unlike existing methods, ALBEF does not require bounding box annotations or high-resolution images and employs momentum distillation to improve learning from noisy network data. 

Moreover, the BLIP model \cite{li2022blip} proposes a new vision-language pre-training framework that is flexible for both understanding and generation tasks. By introducing a boOTStrapped generation and filtering mechanism, BLIP effectively utilizes noisy image-text data from the web. BLIP achieves state-of-the-art performance on various vision-language tasks and demonstrates strong generalization ability, capable of direct zero-shot transfer to video-language tasks. In \cite{yu2022coca}, the Contrastive Captioner (CoCa) model is proposed, which jointly pre-trains an image-text encoder-decoder foundation model with contrastive loss and captioning loss. CoCa effectively combines the advantages of contrastive learning and generative methods and enhances the efficiency of unimodal and multimodal representation learning through a no-cross-attention mechanism.  

\subsection{Advancements in LLMs and Knowledge Distillation Techniques}

The success of GPT-3.5 demonstrated the immense potential and practical value of large language models, sparking extensive research and further development in the field, leading to the emergence of various LLMs. For instance, LLaMA \cite{touvron2023llama}, developed by Meta's research team, showcases excellent text generation and comprehension abilities by training on larger datasets than usual, using only publicly available data, thus promoting the accessibility and research of LLMs. The subsequent LLaMA-2 \cite{touvron2023llama} introduced multiple improvements over LLaMA 1, such as training on a new mixture of publicly available data. 

ChatGLM \cite{du2021glm}, proposed by Tsinghua University, is a conversational model developed under the General Language Model (GLM) framework. It performs excellently in natural language understanding (NLU) and text generation tasks through autoregressive blank filling pre-training. ChatGLM can handle multi-label fill-in-the-blank problems and demonstrates outstanding performance in NLU, conditional generation, and unconditional generation tasks.  Among smaller-scale LLMs, Microsoft's Phi-1.5 model \cite{li2023textbooks} matches the performance of models five times its size on natural language tasks and outperforms most non-frontier LLMs on more complex reasoning tasks such as elementary mathematics and basic coding. In addition to Transformer-based LLMs, non-Transformer LLMs are also being developed. The RWKV model \cite{peng2023rwkv} combines the advantages of RNNs and Transformers while overcoming their key drawbacks. It replaces traditional dot-product attention with linear attention mechanisms, significantly reducing computational and memory complexity. The Mamba model \cite{gu2023mamba} is a new type of selective state-space model (SSMs) \cite{patro2024mamba}, designed with a simple and unified architecture that combines previous SSM architectures with Transformer MLP blocks. 

Despite the rapid development of LLMs, their computational demands and data requirements remain very high, making them difficult for ordinary developers and researchers to utilize directly. Therefore, methods such as Low-Rank Adaptation (LoRA) \cite{hu2021lora} or knowledge distillation are needed to achieve efficient training and deployment of models with reduced computational resources. Knowledge distillation transfers the knowledge of LLMs to smaller models, significantly reducing computational resource and storage needs while maintaining high performance. This technique not only lowers the cost of model deployment but also enhances the practical usability of models in resource-constrained environments, enabling a wider range of developers and researchers to leverage the powerful capabilities of LLMs for innovation and research. Key methods of LLM knowledge distillation include Supervised Fine-tuning \cite{xu2023wizardlm,wang2022self}, Divergence and Similarity \cite{sanh2019distilbert,gu2023minillm,liang2020mixkd,timiryasov2023baby}, Reinforcement Learning \cite{cui2023ultrafeedback,kwon2023reward}, and Rank Optimization \cite{tunstall2023zephyr,hong2023cyclealign}. A detailed discussion of the field of LLM knowledge distillation is provided in \cite{xu2024survey}, and relevant details can be found in that article.

\subsection{Neural Symbolic Regression}
Neural Symbolic Regression (NSR) is a method that combines deep learning and symbolic regression, aiming to automatically discover mathematical expressions that describe the underlying patterns in data. Unlike traditional numerical regression methods, symbolic regression not only provides a model to fit the data but also generates interpretable symbolic expressions. The field of DSR is rapidly evolving, with reinforcement learning and End-to-End (E2E) approaches receiving the most attention.

In the milestone work of \cite{petersen2019deep}, Deep Symbolic Regression (DSR) applies reinforcement learning to optimize the search and generation process of symbolic expressions. In the environment set by the reinforcement learning algorithm, an agent constructs expressions by selecting symbols and operators and receives rewards based on how well these expressions fit the data. Several works have discussed the application of reinforcement learning in DSR and developed new ideas. For example, RL-GEP \cite{zhang2021rl} combines reinforcement learning and genetic algorithms, leveraging the strengths of both to improve the performance in solving symbolic regression problems. Experimental results show that RL-GEP performs excellently on ten benchmark datasets, outperforming methods that use reinforcement learning or genetic algorithms alone. In \cite{crochepierre2022interactive}, the authors propose an interactive reinforcement learning platform for grammar-guided symbolic regression, improving SR results by learning user preferences for expression pairs. 

On the other hand, the End-to-End (E2E) approach uses Transformer models to directly predict solutions for symbolic regression on synthetic datasets. By using a mixed symbolic-numerical vocabulary, symbolic tokens represent operators and variables, while numerical tokens represent constants. A representative work of the E2E method is \cite{kamienny2022end}, where the proposed E2E approach significantly narrows the accuracy gap with state-of-the-art GP techniques in the SRBench benchmark, provides orders-of-magnitude acceleration in inference time, and demonstrates robustness to noise and extrapolation capabilities. Additionally, SymFormer \cite{vastl2024symformer} is another type of E2E method that generates constants alongside symbols, improving model accuracy. The generated constants are used to initialize a local gradient optimizer to fine-tune the final constant values. This method has been comprehensively evaluated on a large number of univariate and bivariate functions and compared with relevant alternative methods.

Compared to reinforcement learning and evolutionary algorithms, E2E methods transform the SR problem into a sequence generation problem, eliminating the need for iterative validation processes, thus being more efficient and gaining rapid development in recent years \cite{shojaee2024transformer,biggio2020seq2seq,arechiga2021accelerating}. However, E2E methods face issues such as not being able to accurately validate and adjust the fitting formulas like evolutionary and reinforcement learning algorithms, which need further research to resolve.

\section{Methodology: Description of Botfip-LLM Framework}

In this chapter, we detail the Botfip-LLM enhanced extension framework, including its model architecture, pre-training process, fine-tuning tasks, and various specifics. The overall framework and pre-training phase flow can be referenced in Figure \ref{fig:Botfip_total_LLM_framework}. For foundational content on the Botfip framework, including the Funcimg-OTS data generation method and dataset format, refer to \cite{chen2024boOTStrapping}.

\subsection{Framework Architecture}

In this section, we provide a detailed overview of the Botfip-LLM architecture, which includes the Funcimg encoder, OTS encoder, OTS decoder, LLM, and its embedder. The components of the Funcimg encoder, OTS encoder, and decoder have been previously described in \cite{chen2024boOTStrapping} with minimal modifications in the Botfip-LLM framework, thus we provide a brief introduction here. Let \(f_{o,c} \in C(\mathbbm{R}^n; \mathbbm{R})\) denote the computable symbolic operation tree corresponding to OTS \(o \in \mathcal{N}_v^k\) and constant vector \(c \in \mathbbm{R}^{d_c}\), where \(\mathcal{N}_v=\{1,...,N_v\},N_v>0\) and \(N_v, d_c > 0\) are the vocab number and constant dimension, respectively. Define the multi-scale meshgrid \(M_\delta \in \mathbbm{R}^{n_s \times d \times n_\delta}\), where \(n_s, n_\delta > 0\) represent the number of multi-scale channels and the number of grid points per dimension, respectively. The function image corresponding to \(f_{o,c}\) can be expressed as \(x_{o,c}^i = f_{o,c}(M_\delta)\). Additionally, define \(x_{o,c}^s = Sym(f_{o,c})\) as the symbolic expression obtained from the function \(f_{o,c}\) via symbolic computation, where \(Sym\) is the symbolic computation operator. In this framework, the dimension of features obtained by all encoders is unified as \(d_f > 0\) for alignment training. Note that the transformer-based encoder may or may not use the cross-attention mechanism to integrate external data features. Let the mapping represented by the model network be \(e\), and we use \(e(\cdot|\cdot)\) to indicate the presence or absence of external condition inputs. When there are no external condition inputs, \(e(\cdot) = e(\cdot|\emptyset)\) denotes the encoder using self-attention, with \(\emptyset\) indicating an empty condition input. When there are external condition features \(h\), \(e(\cdot|h)\) indicates the use of cross-attention.

\textbf{Funcimg Encoder:} The Funcimg encoder primarily handles feature extraction from function images \(x_{o,c}^i\). In the Botfip-LLM framework, we typically choose the ViT model or its variants as the image feature extraction model. Define the expression function corresponding to the Funcimg encoder as \(e^i: \mathbbm{R}^{n_s \times d \times n_\delta} \to \mathbbm{R}^{n_t^i \times d_f}\), where \(n_t^i > 0\) is the number of tokens in the features obtained by the Funcimg encoder. Therefore, the features obtained by the Funcimg encoder from the input \(x_{o,c}^i\) can be expressed as \(h^i = e^i(x_{o,c}^i) \in \mathbbm{R}^{n_t^i \times d_f}\).

\textbf{OTS Encoder/Decoder:} The OTS encoder primarily handles feature extraction from OTS sequences and is used in conjunction with the Funcimg encoder for function image recognition and classification. The OTS decoder is mainly used for subsequent generation tasks. In the Botfip-LLM framework, we choose the BERT model as the backbone architecture. Define the expression functions corresponding to the OTS encoder and decoder backbone networks as \(e^o(\cdot|\cdot), d^o(\cdot|\cdot): \mathcal{N}_v^{\Tilde{k}} \times \mathbbm{R}^{\Tilde{d}_c} \to \mathbbm{R}^{n_t^o \times d_f}\) (regardless of the presence of external condition inputs), where \(n_t^o = \Tilde{k} + \Tilde{d}_c > 0\) is the number of tokens in the features obtained by the Funcimg encoder, \(\Tilde{k}\) and \(\Tilde{d}_c\) are the maximum permissible OTS length and constant length, respectively. If the length of the input OTS and constants is less than these values, they are padded to this length; otherwise, they are clipped\footnote{In general, avoid clipping the OTS due to exceeding the maximum allowable length of the model, as this can result in loss of OTS information, affecting alignment.}. In some tasks, the encoder and decoder may mask parts of the input (especially the constant vector). When there are no external condition inputs, the features of the OTS and constant vector obtained by the OTS encoder can be expressed as \(h^o = h^o(\emptyset) = e^o(o, c) \in \mathbbm{R}^{n_t^o \times d_f}\), and with external features \(h\), as \(h^o(h) = e^o(o, c|h)\). The OTS encoder has a corresponding classification head for related classification task training, which generally uses only the global information of the features (i.e., the hidden states of the first token) or their mean. Therefore, the mapping corresponding to this head can be expressed as \(l^o_e: \mathbbm{R}^{d_f} \to I^{k_o}\), where \(I = (0,1)\) and \(k_o > 0\) is the number of classes, typically 2. On the other hand, the OTS decoder has a corresponding prediction head \(l^o_d: \mathbbm{R}^{n_t^o \times d_f} \to I^{\Tilde{k} \times N_v}\). Note that the OTS decoder uses a casual-attention mechanism instead of cross-attention and is only used to predict the OTS, not the constants. The constant vector is iteratively updated using the L-BFGS algorithm during the inference phase after reconstructing the symbolic operation tree skeleton, simplifying model complexity and parameter count.

\textbf{Frozen LLM and its Embedder:} The most significant difference in the Botfip-LLM framework compared to the original model is the LLM and its embedder. The pre-trained LLMs are primarily used for feature extraction from the function expression string \(x_{o,c}^s\) corresponding to the function \(f_{o,c}\). Define the mapping corresponding to the LLM as \(e^s\), with the size of the dictionary set introduced by the LLMs as \(N_{M} > 0\). The function expression string \(x_{o,c}^s\) is converted into word embedding vectors \(\Tilde{x}_{o,c}^s \in \mathbbm{R}^{N_m \times D_m}\) through the tokenizer matched by the LLMs, where \(N_m, D_m > 0\) are the token number and the embedding dimension of the LLM tokenizer, respectively. The features obtained by the LLM from the function expression string \(x_{o,c}^s\) can be expressed as \(\Tilde{h}^s = e^s(\Tilde{x}_{o,c}^s) \in \mathbbm{R}^{N_m \times D_m'}\), where \(D_m' > 0\) is the feature dimension obtained by the LLM. The dimension of the features \(\Tilde{h}^s\) generally does not match the model-specified feature dimension, requiring the corresponding trainable embedder \(l^s\) to convert \(\Tilde{h}^s\) into \(h^s \in \mathbbm{R}^{N_m \times d_f}\), making it further usable for subsequent pre-training and fine-tuning tasks. In the Botfip-LLM framework, we choose a simple MLP network as the main structure of the LLM embedder. Since the LLMs are frozen and do not participate in training, the corresponding LLM embedder, acting as an adapter, retains only a small number of parameters, significantly reducing training requirements and difficulty. In the following, we will also introduce a trick for the Botfip-LLM framework under distributed training conditions, making it possible to introduce LLMs with low GPU memory.

\subsection{Pre-training and Fine-tuning Process}

In this section, we introduce the pre-training and fine-tuning phases of the Botfip-LLM framework. The pre-training phase of Botfip-LLM primarily involves calculating the Function Image-OTS Contrastive Loss (FOC), Function Image-OTS Matching Loss (FOM), OTS Modeling Loss (OM), and LLM Knowledge Distillation Loss (LLM-KD). To facilitate the calculation of contrastive learning losses, we adopt the concept of queues from the MoCo framework \cite{chen2020improved}, using queues to store historical feature data, thereby reducing the difficulty of negative sample collection and improving computational efficiency. Let the current dataset be represented as \(D_{o,c} = \left\{(o_j,c_j,x^i_{o_j,c_j},x^s_{o_j,c_j})\right\}_{j=1}^{N_d}\), where \(N_d > 0\) is the dataset size. \(Q^i, Q^o, Q^s\) are the corresponding queues for \(h^i, h^o, h^s\) (initially generated by random sampling from white noise and subsequently filled with computed features). First, the loss function for the FOC task \(L_{FOC}\) can be expressed as
\begin{footnotesize}
\begin{align}
\nonumber
    L_{FOC} &= \textbf{InfoNCE}(D_{o,c}) \\
\label{eq:FOC}
    &= - \frac{1}{N_d} \sum_{j=1}^{N_d} \left(\log \frac{\exp(\text{sim}(h^i_j, h^o_j) / \tau)}{\sum_{k=1}^{N_q} \exp(\text{sim}(h^i_j, \overline{h}^o_k) / \tau)} + \log \frac{\exp(\text{sim}(h^o_j, h^i_j) / \tau)}{\sum_{k=1}^{N_q} \exp(\text{sim}(h^o_j, \overline{h}^i_k) / \tau)}\right),
\end{align}
\end{footnotesize}where \(sim\) is the similarity function, typically cosine similarity, \(\tau\) is the temperature coefficient, which is a learnable parameter, and \(\left\{\overline{h}^{i}_{k}, \overline{h}^{o}_{k}\right\}_{k=1}^{N_q}\) are the negative samples of historical features sampled from queues \(Q^i, Q^o\), with \(N_q > 0\) being the sample size. 

The FOM task can be formulated as a binary classification training task, with the loss function \(L_{FOM}\) expressed as
\begin{footnotesize}
\begin{align}
\nonumber
    L_{FOM} = - \frac{1}{N_d} \sum_{j=1}^{N_d} &\left(\sum_{k=1}^{N_{S,+}} \log \frac{\exp(l_e^o(\Tilde{h}^o_j(h^i_{j,k,+}))[0])}{\exp(l_e^o(\Tilde{h}^o_j(h^i_{j,k,+}))[0]) + \exp(l_e^o(\Tilde{h}^o_j(h^i_{j,k,+}))[1])} + \right.\\
    \label{eq:FOM}
    & \left. \sum_{k=1}^{N_{S,-}} \log \frac{\exp(l_e^o(\Tilde{h}^o_j(h^i_{j,k,-}))[1])}{\exp(l_e^o(\Tilde{h}^o_j(h^i_{j,k,-}))[0]) + \exp(l_e^o(\Tilde{h}^o_j(h^i_{j,k,-}))[1])}\right),\\
    \Tilde{h}^o_j(h^i_{j,k,+}) = e^o(o_j, &\Tilde{c}|h^i_{j,k,+}),\ \Tilde{h}^o_j(h^i_{j,k,-})= e^o(o_j, \Tilde{c}|h^i_{j,k,-}),
\end{align}
\end{footnotesize}where \(h^i_{j,k,+}, h^i_{j,k,-}\) represent the features of positive and negative samples of the function image corresponding to OTS \(o_j\) in the training batch, respectively, i.e., whether the function image corresponds to \(o_j\), and \(N_{S,+}, N_{S,-} > 0\) are the numbers of positive and negative samples in the batch, respectively. \(\Tilde{c}\) is the masked constant vector, and \(l_e^o(\Tilde{h}^o_j(h^i_{j,k,+})), l_e^o(\Tilde{h}^o_j(h^i_{j,k,-}))\) are the logits output by the model. 

The OM task, which is the sequence prediction modeling task for OTS, can be expressed as
\begin{footnotesize}
\begin{align}
\label{eq:OM}
        L_{OM} &= -\frac{1}{N_d} \sum_{j=1}^{N_d} \left(\sum_{k=1}^{len(o_j)-1}
    \log \frac{\exp \left( l_d^o(\hat{h}^o_j(h_j^i))[k, o_j[k+1]]\right)}{\sum_{n=1}^{len(o_j)} \exp\left( l_d^o(\hat{h}^o_j(h_j^i))[k, n]\right)}
    \right),\\
    \hat{h}^o_j(h_j^i) & = d^o(o_j, \Tilde{c}|h_j^i),
\end{align}
\end{footnotesize}where \(len(o_j)\) is the length of \(o_j\), and \(o_j[k]\) is the index of the \(k\)-th symbol in the OTS.

\begin{figure}
    \centering
    \includegraphics[width=1.0\textwidth]{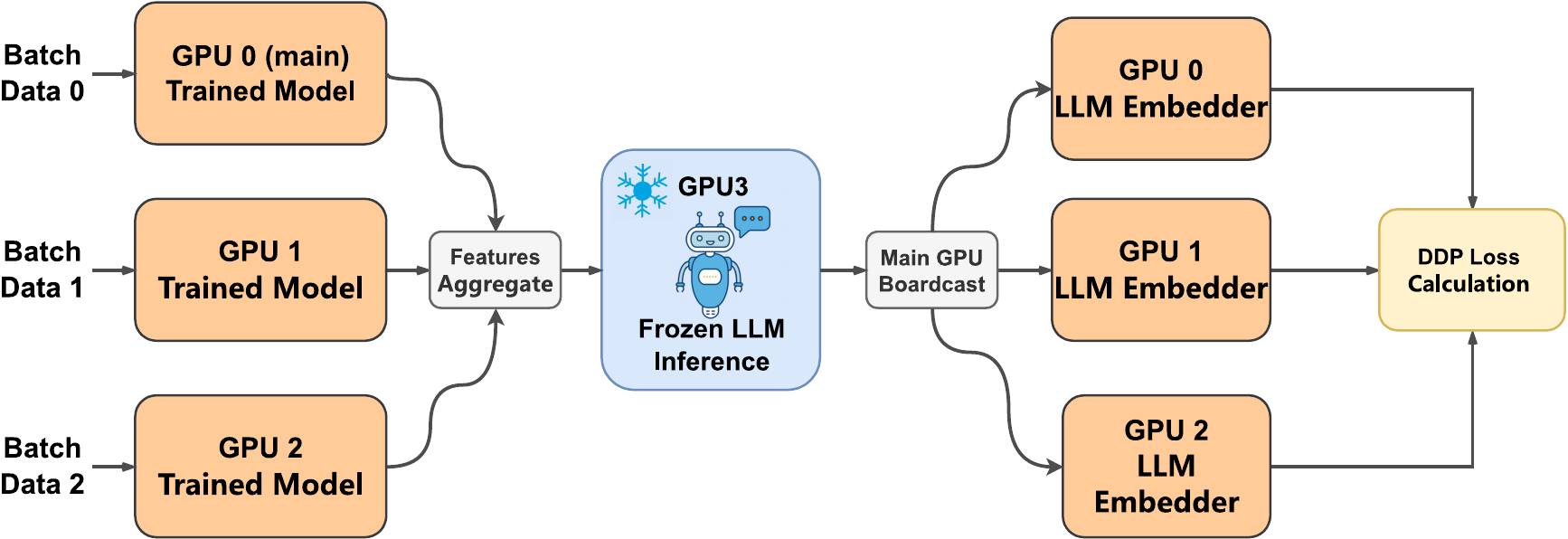}
    \caption{Visualization of the Botfip-LLM distributed training aggregation and broadcast process. The main model is distributed across GPUs 0, 1, and 2, while a half-precision or quantized Frozen LLM model is deployed on GPU 3. During distributed training, features obtained by the models on GPUs 0, 1, and 2 are aggregated and transmitted via the main GPU to the LLM on GPU 3 for feature extraction of the symbolic expression \(x^s_{o,c}\). The results are then broadcasted back to the main models on other GPUs for further computation.}
    \label{fig:DDP-process-cropped}
\end{figure}

Next, we introduce how LLMs guide the OTS encoder and Funcimg encoder through knowledge distillation and contrastive learning during the pre-training phase. In Botfip-LLM, we primarily use the Similarity method in knowledge distillation, i.e., guiding the training of student models (OTS encoder and Funcimg encoder) by comparing the hidden states and features of the teacher model (LLMs) with those of the student model \cite{xu2024survey}. Specifically, we align the features of the LLMs, \(h^s\), with \(h^i\) and \(h^o\) for training. This method plays a crucial role in the Botfip-LLM framework, ensuring that the internal representations of the student model are highly consistent with those of the teacher model. This not only improves the accuracy of the student model in generating outputs but also enhances the consistency of the model in the information processing process. This means that the student model can better mimic the behavior and decision logic of the teacher model when facing complex tasks, thereby significantly improving the model's generalization ability and robustness. At this point, the LLM knowledge distillation task loss \(L_{KD}\) can be expressed as
\begin{footnotesize}
\begin{align}
\label{eq:KD}
    L_{KD} &= - \frac{1}{N_d} \sum_{j=1}^{N_d} \log \frac{\exp(\text{sim}(h^i_j, h^s_j) / \tau)}{\sum_{k=1}^{N_q} \exp(\text{sim}(h^i_j, \overline{h}^s_k) / \tau')} + \log \frac{\exp(\text{sim}(h^o_j, h^s_j) / \tau)}{\sum_{k=1}^{N_q} \exp(\text{sim}(h^o_j,  \overline{h}^s_k) / \tau')},
\end{align}
\end{footnotesize}where \(\tau' > 0\) is also a temperature coefficient, and \(\left\{\overline{h}^s_k\right\}_{k=1}^{N_q}\) are the negative samples of historical symbolic formula encoding features collected from queue \(Q^s\). By using the Similarity method, Botfip-LLM effectively guides the OTS encoder and Funcimg encoder, enabling the student model to efficiently learn and absorb the knowledge of the teacher model. This approach in the Botfip-LLM framework not only helps align the output features of the Funcimg and OTS encoders with the embedded features of the LLMs but also makes the information processing methods more similar, thereby achieving more efficient and precise knowledge transfer. Ultimately, the total loss function for the pre-training phase \(L_{pre}\) is given by
\begin{footnotesize}
\begin{align}
    L_{pre} = \lambda_1 L_{FOC} + \lambda_2 L_{FOM} + \lambda_3 L_{OM} + \lambda_4 L_{KD},
\end{align}
\end{footnotesize}where \(\lambda_1, \lambda_2, \lambda_3, \lambda_4\) are the respective weight coefficients.

\begin{algorithm}
\caption{Botfp-LLM Pre-training Procedure}
\begin{algorithmic}[1]
\STATE \textbf{Input:} Funcimg-OTS-Formula Dataset \(D_{o,c} = \left\{(o_j,c_j,x^i_{o_j,c_j},x^s_{o_j,c_j}\right\}_{j=1}^{N_d}\), OTS Encoder $e^o$ and its classification head \(l_e^o\), OTS Decoder $d^o$ and its prediction head \(l_d^o\), Funcimg Encoder $e^i$, Pre-trained Frozen LLM \(e^s\) and its embedder \(l^s\), Queues $(Q^i,Q^o,Q^s)$, Number of Epochs $E$, Learning Rates $\lambda_1$, $\lambda_2$, $\lambda_3$, $\lambda_4$, Queue Size $Q$, Temperature $\tau,\tau'$, masked constant array \(\Tilde{c}\).
\STATE Initialize queues $Q_I$, $Q_O$ and bind the weights of the \(e^o\) and \(d^o\).
\FOR{$e \leftarrow 1$ \TO $E$}
    \FOR{each batch $\left(o_j, c_j, x^i_{o_j,c_j}, x^s_{o_j,c_j}\right)$ in $D_{o,c}$}
    \STATE \textbf{\#Calculate Features}
        \STATE $h^i_j = e^i_j(x_{o_j,c_j}^i)$
        \STATE $h^o_j = e^o(o_j, c_j)$
        \STATE $\Tilde{h}^o_j(h^i_j) = e^o(o_j, \Tilde{c} \mid h^i_j)$
        \STATE $\hat{d}^o_j(h^i_j) = d^o(o_j, \Tilde{c} \mid h^i_j)$ 
        
        \IF{Using Distributed learning}
            \STATE Aggregate \(x_{o,c}^s\) from different GPUs to the GPU where the LLM is deployed
            \STATE $\Tilde{h}^s = e^s(x_{o,c}^s)$
            \STATE Split \(\Tilde{h}^s\) and broadcast the corresponding features to the respective GPUs
        \ENDIF
        
        \STATE $h^s = l^s(\Tilde{h}^s)$
        
        \STATE \textbf{\#Sample Negative Examples from the Queue}
        \STATE $\left\{\overline{h}^{i}_{k}, \overline{h}^{o}_{k}, \overline{h}^{s}_{k}\right\}_{k=1}^{N_q} \sim (Q^i, Q^o, Q^s)$
        
        \STATE Calculate \(L_{FOC}, L_{FOM}, L_{OM}, L_{KD}\) through eq.(\ref{eq:FOC}), (\ref{eq:FOM}), (\ref{eq:OM}), (\ref{eq:KD})
        \STATE $L_{pre} = \lambda_1 L_{FOC} + \lambda_2 L_{FOM} + \lambda_3 L_{OM} + \lambda_4 L_{KD}$
        \STATE Update trained parameters with respect to $L_{pre}$ using the optimizer
        \STATE Update queues $(Q^i, Q^o, Q^s)$ cyclically
    \ENDFOR
\ENDFOR
\end{algorithmic}
\label{alg: Pre-training Procedure}
\end{algorithm}

In the engineering implementation of the pre-training process, we use a queue to store historical data to enhance the effectiveness of contrastive learning. Specifically, we store samples generated in the previous training rounds in a fixed-size queue and compare the current batch of samples with those in the queue during each training round. The main advantage of this method is that it increases the diversity of negative samples, enabling the model to better learn the differences between different modalities. Additionally, during the pre-training phase, we typically share the weights of the OTS encoder \(e^o\) and decoder \(d^o\) to reduce the number of pre-training parameters and accelerate the training process.

Furthermore, in the case of distributed training, handling the loading of LLMs on GPUs becomes crucial, especially under low GPU memory conditions. If a distributed data parallel strategy is employed, LLMs need to be loaded onto all GPUs, which significantly consumes GPU memory and may hinder training. Therefore, we adopt a distributed aggregation approach, where an independent parameter-frozen LLM is deployed on a single GPU, while the Botfip-LLM main model is deployed on other GPUs for distributed training, which is shown in Figure \ref{fig:DDP-process-cropped}. At this point, the deployment of LLMs on GPUs can utilize half-precision or even 8-bit or 4-bit quantization methods to further reduce the model parameters and computational load, thereby improving LLM computational efficiency. When encoder output features are generated on different GPUs, these features are aggregated to the GPU hosting the LLM through inter-GPU data transfer. After inference by the LLM, the resulting features are broadcasted by the main GPU to the corresponding GPUs for distributed training. The process flow is illustrated in figure. This method significantly reduces the computational requirements of the model and avoids redundant deployment of LLMs, enabling multimodal frameworks like Botfip-LLM to perform distributed training under low GPU memory conditions, thereby alleviating the computational burden on future scientific researchers. The pseudocode for the Botfip-LLM pre-training phase can be found in Algorithm \ref{alg: Pre-training Procedure}.

\subsection{Fine-tuning Task:  Formula String-OTS Transformation}

In previous fine-tuning tasks, such as the Funcimg-OTS generation fine-tuning task, details can be found in \cite{chen2024boOTStrapping}. The original Botfip model achieved the recognition of function images by fine-tuning the OTS decoder, thereby generating the corresponding OTS. However, we previously lacked the ability to directly transform symbolic formula strings into OTS. This might seem straightforward since we could manually calculate the transformation from symbolic expressions to OTS, but in reality, it presents many challenges. The more complex the symbolic expression, the harder it is to transform it into the corresponding symbolic operation tree and thus obtain the OTS. There are numerous calculations that might involve simplifications and other processes, complicating the transformation from the original symbolic operation tree. First, the complexity of symbolic formulas brings about structural parsing difficulties. Complex formulas contain multi-layered nested operations and various mathematical symbols, requiring precise parsing of their syntax and semantic structure. Second, converting formulas into symbolic operation trees involves considering the precedence and associativity of different operators, increasing the complexity of the transformation. Moreover, the diversity of variables and parameters in symbolic formulas demands that the model has a high degree of generalization capability to handle various forms of input. Another notable issue is that different structures of symbolic operation trees might correspond to the same symbolic formula, implying that the relationship between symbolic formulas and symbolic operation trees is not one-to-one. Thus, obtaining a simplified OTS from symbolic formulas is also a challenge.

Thanks to the introduction of LLM and its embedder, we were able to extend the original Botfip model, adding the capability to transform symbolic formula strings into OTS. In this fine-tuning task, similar to the pre-training phase, the LLM and its embedder extract features from the input  formula string \(x^s_{o,c}\), generating feature representations \(h_s^i\) of the symbolic formula. These feature representations capture the syntactic and semantic information of the formula, providing enough context to understand the structure and content of the formula. The extracted feature representations \(h_s^i\) are then passed to the OTS decoder through a Causal-Attention mechanism. The Causal-Attention mechanism ensures that the sequence dependency is maintained during the generation process, resulting in coherent and accurate OTS. The loss function for the corresponding fine-tuning task, \(L_{SOM}\), can be expressed as follows:
\begin{align}
    L_{SOM} &= -\frac{1}{N_d} \sum_{j=1}^{N_d} \left(\sum_{k=1}^{\text{len}(o_j)-1}
    \log \frac{\exp \left( l_d^o(\hat{h}^o_j(h_j^s))[k,o_j[k+1]]\right)}{\sum_{n=1}^{\text{len}(o_j)} \exp\left( l_d^o(\hat{h}^o_j(h_j^s))[k,n]\right)}
    \right),\\
    \hat{h}^o_j(h_j^s) & = d^o(o_j, \Tilde{c} \mid h_j^s).
\end{align}

Finally, the OTS decoder generates the corresponding optimized transfer sequence based on the input feature representations. Through this process, we achieve the automatic transformation from symbolic formula strings to OTS, avoiding the cumbersome steps of manual calculation and improving the accuracy and efficiency of the transformation. This method not only addresses the difficulties posed by the complexity of symbolic formulas but also provides the model with greater flexibility and scalability, significantly simplifying the process of handling symbolic expressions.

\section{Experiments}

In this chapter, we present the experiments, training, and validation results of Botfip-LLM. The training and validation datasets used in this study are consistent with the open-source Funcimg-OTS dataset used in Botfip. Details on the dataset, main model, and relevant hyperparameters during training can be found in Table \ref{tab: Hyper-parameters of the Dataset and Experiment}. Some examples of the Funcimg-OTS-Formula String multimodal dataset can be found in Appendix \ref{sec: Appendix of Display Selected Skeleton Information}. Similarly, we employed three evaluation metrics: the regularity of OTS generation \(Acc_{r}\), relative sequence Levenshtein similarity \(S_{RL}\), and relative formula-string Levenshtein similarity \(\Tilde{S}_{RL}\), defined as follows:
\begin{align}
    Acc_{r}\left(\Omega \right) &= \frac{1}{N} \sum_{i=1}^N \mathbbm{1}\left( A_i \right),\\
    S_{RL}\left(\Omega\right) &= \frac{1}{N} \sum_{i=1}^N \frac{l(t_i) - D_{RL}\left(p_i,t_i\right)}{l(t_i)},\\
    \Tilde{S}_{RL}\left(\Omega\right) &= \frac{1}{N} \sum_{i=1}^N \frac{\mathbbm{1}\left( A_i \right)\left(l(\Tilde{t}_i) - D_{RL}\left(\Tilde{p}_i,\Tilde{t}_i\right)\right)}{l(\Tilde{t}_i)},
\end{align}
where \(A_i\) denotes the event where \(p_i\) can be reconstructed as an operation tree, \(D_{RL}\) is the Levenshtein distance \cite{yujian2007normalized}, \(l(t_i)\) is the length of the operation tree sequence \(t_j\), and \(\Tilde{p}_i\) and \(\Tilde{t}_i\) are the formula string expressions obtained after reconstructing the operation tree sequence into the operation tree. \(Acc_{r}\) measures the degree to which the generated operation tree sequences can be restored into the operation tree structure, serving as a metric for sequence regularity. \(S_{RL}\) and \(\Tilde{S}_{RL}\) not only incorporate \(Acc_{r}\) but also reflect the differences between the target and predicted symbolic expressions after being reduced to operation trees and computed. These metrics validate the effectiveness of OTS generation. 
\begin{table}[!t]
\centering
\caption{Introduced Pre-trained LLMs Information}
\label{tab: Introduced Pre-trained LLMs Information}
\begin{tabular}{cccc}
\toprule[0.8pt]
\textbf{LLM Name} & \textbf{HuggingFace Path} & \textbf{Params} & \textbf{Hidden Size}\\ \hline
LLaMA-2 & meta-llama/Llama-2-7b & 7B & 32000\\
ChatGLM-2 & THUDM/chatglm2-6b & 6B & 65024\\
Gemma &  google/gemma-7b  & 7B&256000\\
Mistral & unsloth/mistral-7b-v0.2 & 7B & 32000\\
Mamba& TRI-ML/mamba-7b-rw&7B& 50432\\
Qwen-1.5& Qwen/Qwen1.5-7B & 7B& 151936\\
RWKV-6& RWKV/rwkv-6-world-7b& 7B & 65536\\
Phi-2& microsoft/phi-2& 2.8B& 51200\\ 
\bottomrule[0.8pt]
\end{tabular}

\end{table}
\begin{figure}[!t]
    \centering
    \includegraphics[width=1.0\textwidth]{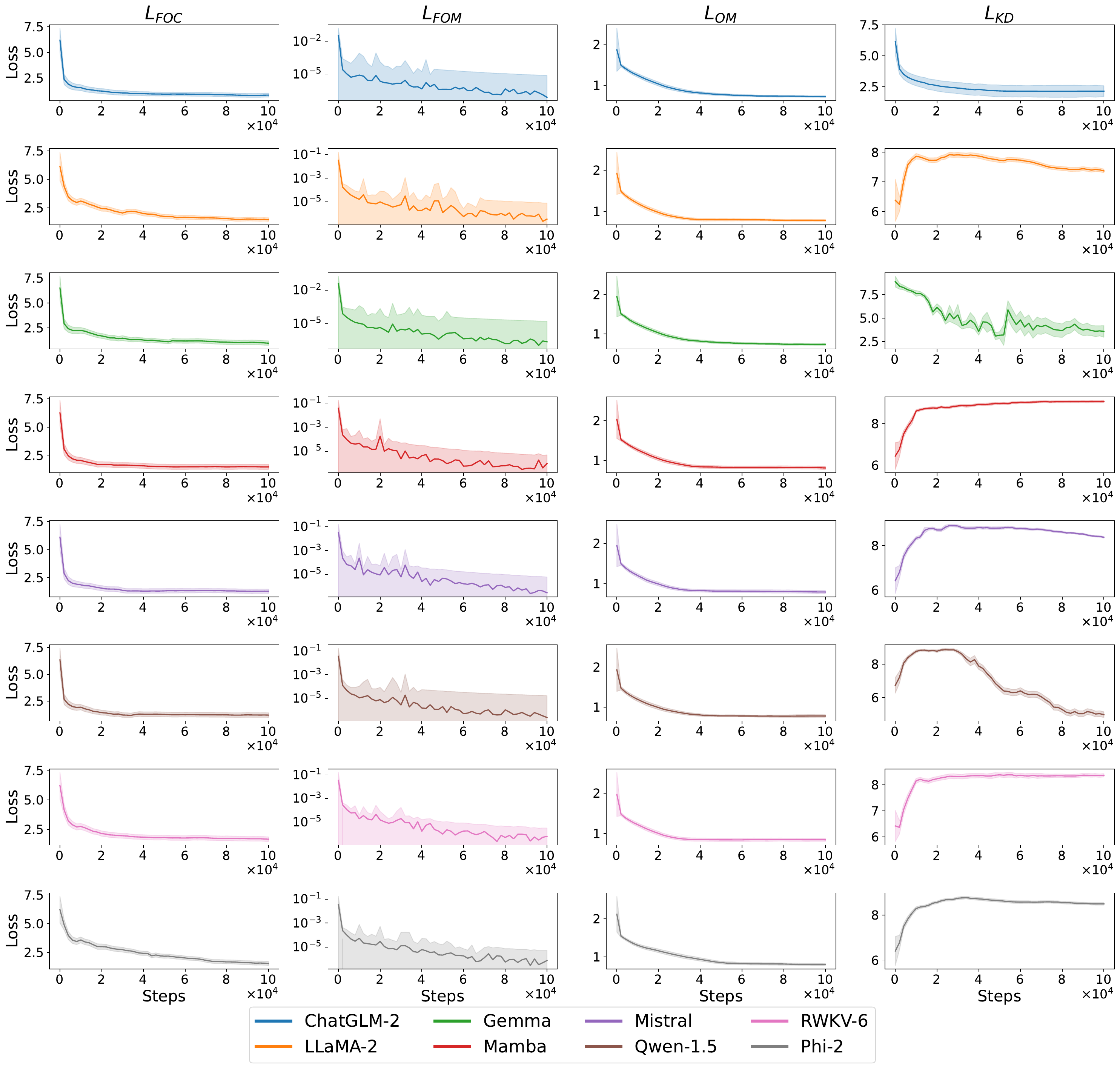}
\caption{During the pre-training phase, the mean and standard deviation of the loss functions \(L_{FOC}\), \(L_{FOM}\), \(L_{OM}\), and \(L_{KD}\) are visualized for the Botfip-LLM framework with the introduction of eight different LLMs. The y-axis of the \(L_{FOM}\) plot uses a logarithmic scale, while the y-axes of the other plOTS use a linear scale.}
    \label{fig:pretrain_loss_curve}
\end{figure}

\subsection{Pre-training Performance with Knowledge Distillation from LLMs}

In this section, we will showcase the performance of Botfip-LLM during the pre-training phase. A crucial question is which LLMs can achieve the best knowledge distillation performance within the Botfip-LLM multimodal scientific computing framework, thereby aiding Botfip-LLM in better multimodal information mining of function sets generated by symbolic operation trees. Here, we have selected 8 LLM models to serve as the teacher model and the main network for symbolic formula feature extraction in Botfip-LLM: LLaMA-2 \cite{touvron2023llama}, ChatGLM-2 \cite{zeng2022glm}, Gemma \cite{team2024gemma}, Mistral \cite{jiang2023mistral}, Mamba \cite{gu2023mamba}, Qwen-1.5 \cite{bai2023qwen}, RWKV-6 \cite{peng2023rwkv}, and Phi-2 \cite{li2023textbooks}. It is important to note that Mistral’s official weights require an API key, so we have chosen alternative weights for this model, while the weights for the other LLMs are the official weights provided by their respective institutions. Additionally, Phi-2 is a smaller LLM with a parameter count of up to 2.8B, whereas the other models range from 6B to 7B parameters. Specific information about the introduced LLMs, including their pre-trained model checkpoints and HuggingFace paths, as well as the hidden sizes of different LLM networks, can be found in Table \ref{tab: Introduced Pre-trained LLMs Information}. It should be noted that the pre-training corpus has a significant impact on the pre-trained LLM models. Therefore, the results of knowledge distillation training in this paper cannot solely reflect the advantages or disadvantages of the LLMs' architectures in Botfip-LLM knowledge distillation pre-training. We aim to identify the most suitable pre-trained LLM model weights for Botfip-LLM through this comparative approach.

\begin{table}[!t]
\centering
\caption{Average training error and its standard deviation at the \num{1e6}th step during the pre-training phase of Botfip-LLM under different LLMs as teacher models.}
\label{tab: pretrain loss}
\resizebox{\textwidth}{!}{
\begin{tabular}{ccccc}
\toprule[0.8pt]
\textbf{LLM Name} & \textbf{\(L_{FOC}\)} & \textbf{\(L_{FOM}\)}& \textbf{\(L_{OM}\)} & \textbf{\(L_{KD}\)}\\ \hline
LLaMA-2 &\(1.16\pm0.16\)&\(\num{3.51E-7}\pm\num{7.58E-6}\)&\(0.78\pm0.02\)&\(7.37\pm0.06\)\\
ChatGLM-2 &\(0.82\pm0.19\)&\(\num{7.66E-8}\pm\num{7.43E-6}\)&\(0.68\pm0.02\)&\(2.15\pm0.44\)\\
Gemma &\(0.97\pm0.23\)&\(\num{2.43E-7}\pm\num{1.62E-5}\)&\(0.72\pm0.03\)&\(3.57\pm0.62\)\\
Mistral &\(1.29\pm0.19\)&\(\num{2.77E-7}\pm\num{5.62E-6}\)&\(0.79\pm0.04\)&\(8.08\pm0.04\)\\
Mamba &\(1.47\pm0.22\)&\(\num{9.33E-7}\pm\num{4.02E-6}\)&\(0.82\pm0.04\)&\(9.08\pm0.05\)\\
Qwen-1.5 &\(0.90\pm0.22\)&\(\num{2.42E-7}\pm\num{1.68E-5}\)&\(0.71\pm0.03\)&\(3.98\pm0.17\)\\
RWKV-6 &\(1.66\pm0.20\)&\(\num{5.94E-7}\pm\num{2.48E-6}\)&\(0.84\pm0.04\)&\(9.36\pm0.06\)\\
Phi-2 &\(1.34\pm0.17\)&\(\num{7.62E-7}\pm\num{4.45E-6}\)&\(0.80\pm0.03\)&\(8.50\pm0.05\)\\
\bottomrule[0.8pt]
\end{tabular}
}
\end{table}

\begin{figure}[!t]
    \centering
    \subfloat[ChatGLM-2 (Before Pre-training)]{
        \includegraphics[width=0.8\textwidth]{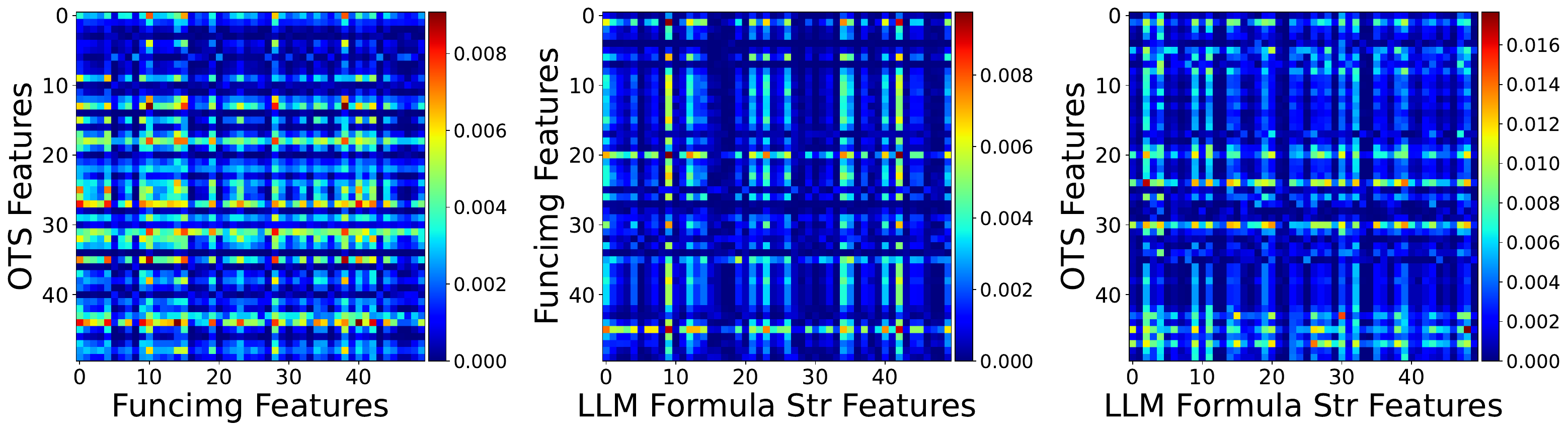}
        \label{fig:contrast ChatGLM-2(Before Pre-training)}
    }\\ 
    \subfloat[ChatGLM-2 (After Pre-training)]{
        \includegraphics[width=0.8\textwidth]{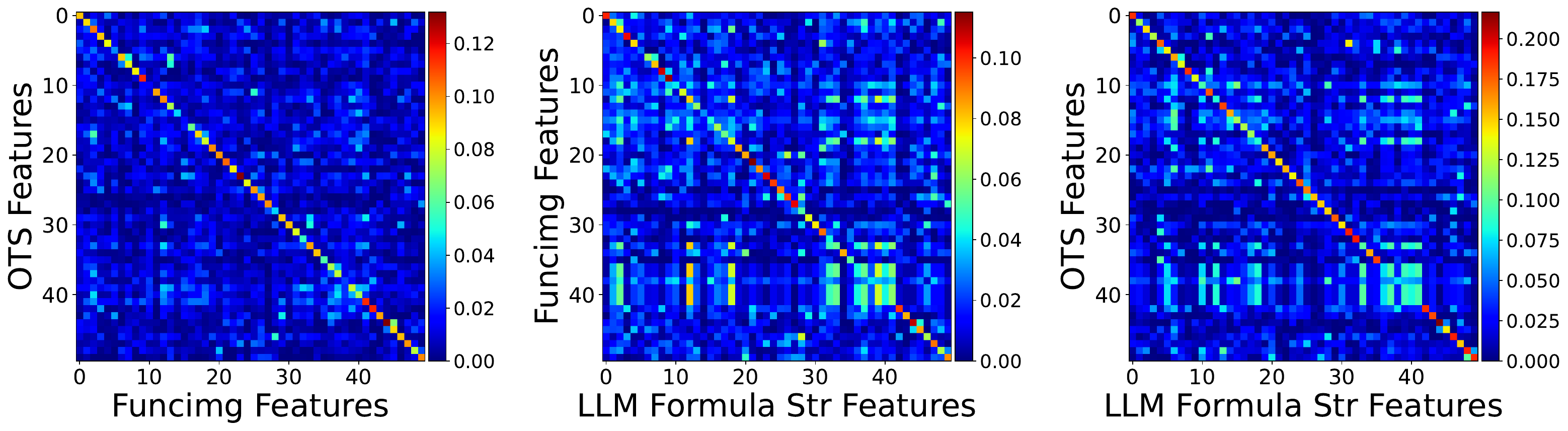}
        \label{fig:contrast ChatGLM-2(After Pre-training)}
    }\\ 
    \subfloat[RWKV-6 (After Pre-training)]{
        \includegraphics[width=0.8\textwidth]{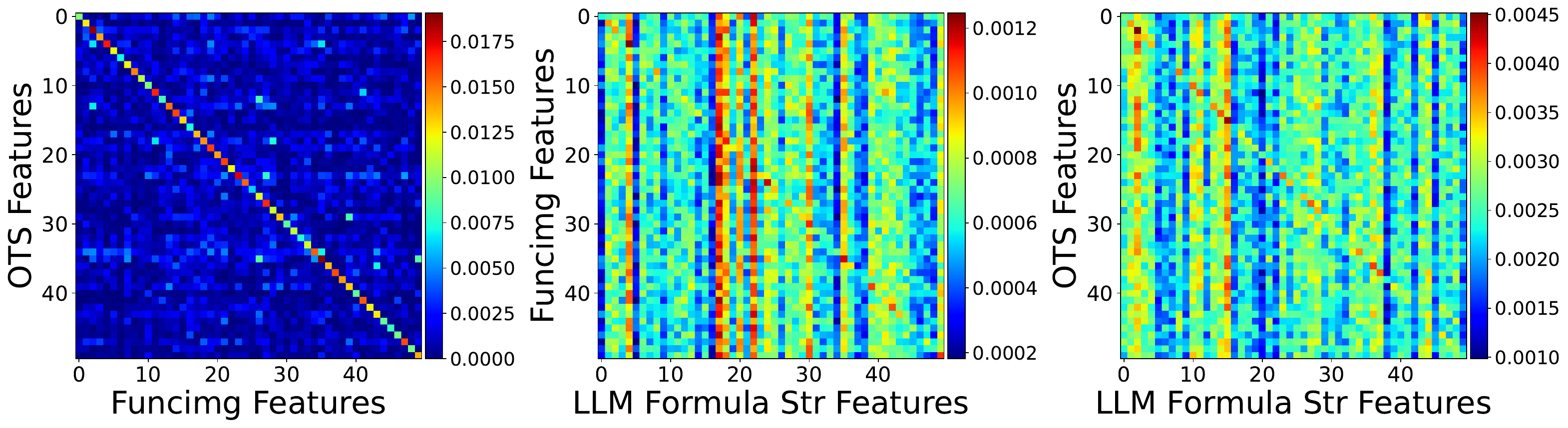}
        \label{fig:contrast RWKV-6(After Pre-training)}
    }
    \caption{Visualization of the cosine similarity matrices between Funcimg features, OTS features, and Formula Str features generated by different encoders in the Botfip-LLM framework for 50 validation samples before and after pre-training. Figures \ref{fig:contrast ChatGLM-2(Before Pre-training)} and \ref{fig:contrast ChatGLM-2(After Pre-training)} show the results with ChatGLM-2 as the Teacher LLM, while Figure \ref{fig:contrast RWKV-6(After Pre-training)} presents the results with RWKV-6 as the Teacher LLM after pre-training.}
    \label{fig:contrast vis}
\end{figure}

Figure \ref{fig:pretrain_loss_curve} shows the loss curves for the four different loss functions \(L_{FOC}\), \( L_{FOM}\), \( L_{OM}\) and \(L_{KD}\)  over \num{1e6} steps during the pre-training phase for the Botfip-LLM framework with eight different LLMs introduced as the Teacher models for the symbolic formula feature extraction network. Additionally, Table \ref{tab: pretrain loss} presents the specific average training error and its standard deviation at the \num{1e6}th step. From the trend of the four loss functions in Figure \ref{fig:pretrain_loss_curve}, it can be observed that while \(L_{FOC}, L_{FOM}, L_{OM}\) all show a decreasing trend under the guidance of different LLMs, \(L_{KD}\) varies greatly with the type of LLM. When ChatGLM-2, Gemma, and Qwen-1.5 are used as Teacher models, \(L_{KD}\) shows a significant downward trend, with ChatGLM-2 decreasing the fastest and most stably, Gemma showing considerable fluctuation, and Qwen-1.5 initially rising and then showing a marked decrease. Conversely, LLaMA-2, Mistral, and Phi-2 exhibit a rapid initial increase in \(L_{KD}\), followed by a slow decline after approximately \num{4e4} steps, with LLaMA-2 decreasing the fastest, followed by Mistral, and Phi-2 the slowest. Lastly, Mamba and RWKV-6 show a rapid increase in \(L_{KD}\) which barely decreases within \num{1e6} steps. For these non-Transformer LLMs, the differences in model architecture likely cause significant differences in the extraction of multimodal information features, making it challenging to effectively transfer knowledge using standard similarity-based distillation methods. For Transformer-based LLMs like ChatGLM-2, Gemma, and Qwen-1.5, we believe that the pre-training corpus and the hidden size of the final model features have a more significant impact on the knowledge distillation training curves shown in Figure \ref{fig:pretrain_loss_curve} when the LLM parameters are of the same magnitude. Both ChatGLM-2 and Qwen-1.5 are large Chinese models extensively using Chinese corpora during pre-training, possibly including a large amount of mathematical-related corpus, and both support 8K-32K contexts. It can be seen that ChatGLM-2 quickly and stably reduces the knowledge distillation loss during the pre-training phase, while Qwen-1.5, although slightly inferior to ChatGLM-2, also shows a rapid decrease in \(L_{KD}\) after a certain training stage. The difference between these two is that Qwen-1.5 has a hidden size of 151936, while ChatGLM-2 has a hidden size of 65024. A larger hidden size means that the LLM embedder requires more time for parameter tuning, and the closer the hidden size is to the student model, the more similar the LLM feature representation is to the standard BERT model, making adjustments easier. On the other hand, the Gemma model, which does not use a multilingual corpus, shows a \(L_{KD}\) loss that, despite some fluctuations, continues to decline. This could be because the Gemma model's pre-training corpus contains a large amount of math-related elements, and it uses multi-query attention mechanisms and Rotary Positional Embedding (RoPE) for enhanced general text understanding capabilities. Lastly, Phi-2 shows the slowest \(L_{KD}\) loss reduction among Transformer-based LLMs, possibly because Phi-2 has a smaller parameter size and its ability to extract features from mathematical symbols is inferior to other Transformer-based LLMs.

From Figure \ref{fig:pretrain_loss_curve} and Table \ref{tab: pretrain loss}, it can be seen that ChatGLM-2 shows relatively the best performance during the pre-training phase. Now we proceed with further testing on the pre-training validation dataset. We randomly sample 50 multimodal samples from the validation dataset and use the corresponding encoders to obtain the respective Funcimg features, OTS features, and Formula Str features generated by the LLM and its embedder. We then compute the cosine similarity between the features of different samples and visualize them in Figure \ref{fig:contrast vis}, including the cases of ChatGLM-2 and RWKV-6 as the Teacher LLM. It is evident that the alignment between different modalities before and after pre-training with ChatGLM-2 as the Teacher LLM is very distinct. The feature extraction capability for the function symbol formula of ChatGLM-2 is effectively transferred to the student model of Botfip-LLM, i.e., the Funcimg Encoder and OTS Encoder, through similarity knowledge distillation. On the other hand, although the Funcimg Encoder and OTS Encoder of Botfip-LLM achieve self-alignment through pre-training with RWKV-6 as the Teacher LLM, the features obtained by these encoders do not successfully align with the character features obtained by RWKV-6, indicating that RWKV-6's knowledge was not successfully transferred to the main model of Botfip-LLM, forming a stark contrast with ChatGLM-2 and other Transformer LLMs.

\subsection{Fine-tuning Performance of Botfip-LLM under LLMs}

In this chapter, we focus on fine-tuning experiments of Botfip-LLM in different downstream tasks, including the Funcimg-OTS Modeling fine-tuning task and the LLM Formula String-OTS fine-tuning task. The former has already been experimented and analyzed within the Botfip framework, while the latter is a new fine-tuning task enabled by the introduction of the LLM model within the Botfip-LLM framework. This expansion allows the LLM to not only serve as a Teacher Model for knowledge distillation but also to broaden the range of downstream tasks in this multimodal scientific computing framework by incorporating LLM components. In the fine-tuning tasks, we mainly test against the original Botfip framework and Botfip-LLM using ChatGLM-2 and RWKV-6 as the Teacher LLM and symbolic formula feature extraction network. This is because the performance of these two models in the pre-training phase is representative, and the pre-training performance of other models generally falls between these two. Therefore, we select the Botfip-LLM pre-trained models corresponding to these two LLMs for the fine-tuning phase experiments.

\subsubsection{Funcimg-OTS Modeling Task Performance}
\begin{figure}[!t]
    \centering
    \includegraphics[width=1.0\linewidth]{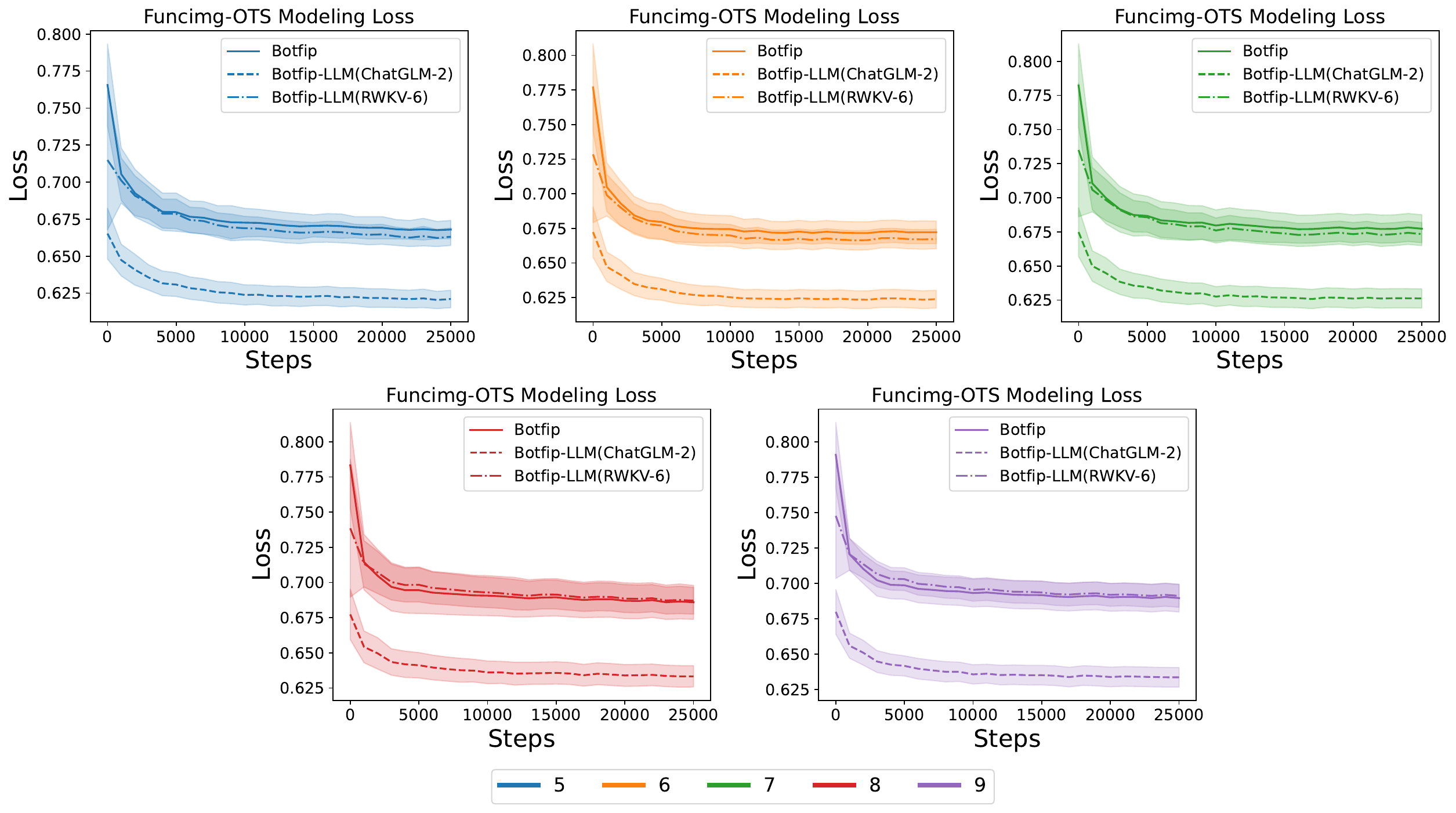}
    \caption{Training loss curves for the Funcimg-OTS Modeling fine-tuning task with different numbers of nodes in the symbolic operation tree. The curves represent the training progress for the original Botfip model, Botfip-LLM (with ChatGLM-2 as the Teacher LLM), and Botfip-LLM (with RWKV-6 as the Teacher LLM).
}
    \label{fig:funcimg_OTS_modeling}
\end{figure}

Now, we proceed with the fine-tuning test of the Funcimg-OTS Modeling task. In this fine-tuning task, we will evaluate the fine-tuning performance of the Botfip-LLM model under the pre-training conditions of different LLMs as Teacher Models and validate the results on the validation dataset. Following the fine-tuning process in \cite{chen2024boOTStrapping}, we will select multimodal datasets with different numbers of nodes as the fine-tuning dataset. This selection is crucial because the number of nodes in the computation tree determines the length of the OTS, thereby reflecting the difficulty of the OTS Modeling fine-tuning task.

\begin{table}[!ht]
\centering
\caption{Mean fine-tuning error and its standard deviation at the \num{25000}th step of Botfip and Botfip-LLM under ChatGLM-2 and RWKV-6 in Funcimg-OTS Modeling Task.}
\label{tab:Funcimg-OTS Modeling Task loss}
\resizebox{\textwidth}{!}{
\begin{tabular}{cccccc}
\toprule[0.8pt]
\textbf{Model} & node \(5\) & node \(6\) & node \(7\) & node \(8\) & node \(9\)\\ \hline
Botfip & \(0.668\pm0.006\) & \(0.672\pm0.008\) & \(0.677\pm0.010\) & \(0.686\pm0.012\) & \(0.690\pm0.010\)\\
\makecell{Botfip-LLM \\ (ChatGLM-2)} & \(0.621\pm0.006\) & \(0.624\pm0.006\) & \(0.626\pm0.007\) & \(0.633\pm0.008\) & \(0.634\pm0.007\)\\
\makecell{Botfip-LLM \\ (RWKV-6)} & \(0.663\pm0.006\) & \(0.667\pm0.007\) & \(0.673\pm0.008\) & \(0.687\pm0.009\) & \(0.691\pm0.008\)\\
\bottomrule[0.8pt]
\end{tabular}
}
\end{table}

As shown in Figure \ref{fig:funcimg_OTS_modeling}, the loss variations in the Funcimg-OTS Modeling fine-tuning task for the original Botfip model and the Botfip-LLM model with ChatGLM-2 and RWKV-6 as pre-trained Teacher LLMs are presented in detail. It is evident from the figure that the Botfip-LLM models exhibit lower initial errors and faster reduction rates compared to the original Botfip model.

Furthermore, the pre-trained model using ChatGLM-2 as the Teacher LLM shows superior performance in terms of convergence speed and final error compared to the model using RWKV-6. Table \ref{tab:Funcimg-OTS Modeling Task loss} further details the steady-state loss average errors and corresponding standard deviations for the original Botfip model and the Botfip-LLM models in the Funcimg-OTS Modeling fine-tuning task. According to the data in the table, the Botfip-LLM model using ChatGLM-2 as the pre-trained model demonstrates a lower average error and smaller standard deviation in the loss function variation compared to the original Botfip model. On the other hand, when RWKV-6 is used as the Teacher Model during the pre-training phase, its loss variation curve is similar to that of the original Botfip model, showing only slight improvement in the initial training phase. This indicates that during the pre-training process, ChatGLM-2 effectively transfers its knowledge to the Botfip-LLM main model through knowledge distillation, enabling better integration of symbolic formula modality information, thereby improving model performance. In contrast, RWKV-6, under specific weights, fails to effectively guide the Botfip-LLM main model during pre-training, resulting in no significant performance improvement in the fine-tuning phase compared to the original model.
\begin{figure}[!t]
    \centering
    \includegraphics[width=1.0\linewidth]{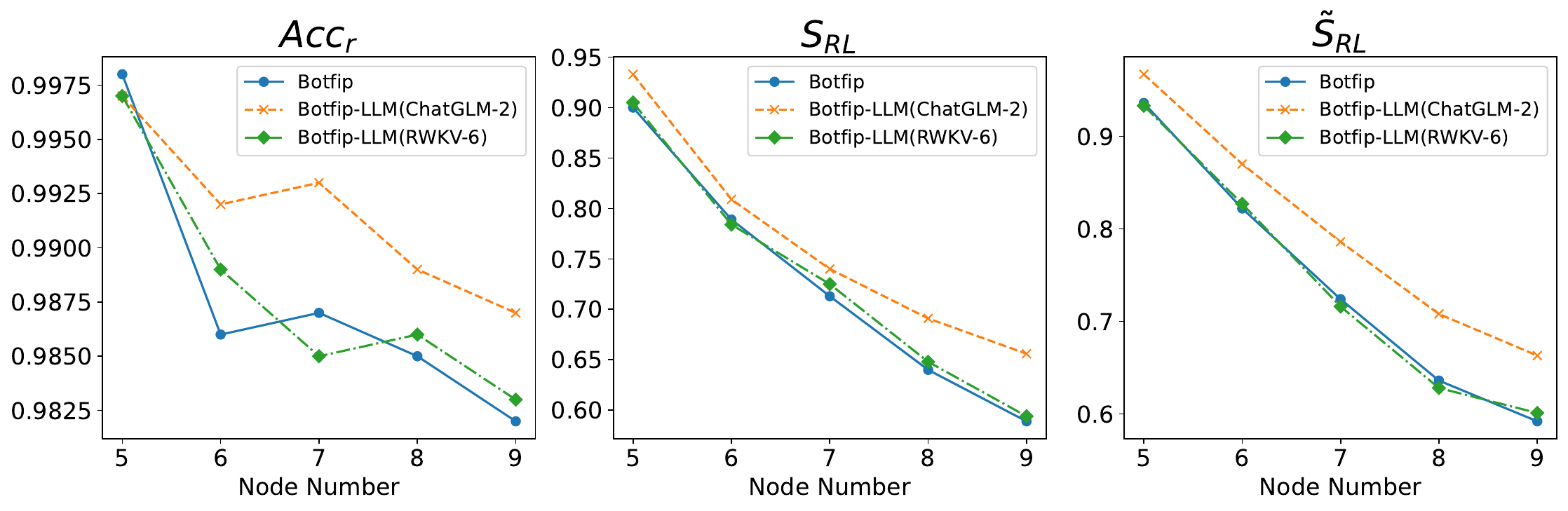}
    \caption{Illustration of the variation in \(Acc_r\), \(S_{RL}\), and \(\Tilde{S}_{RL}\) metrics with respect to the number of nodes in the symbolic operation tree during the Funcimg-OTS Modeling fine-tuning task. The different line styles represent the performance of the original Botfip model, Botfip-LLM with ChatGLM-2 as the Teacher LLM, and Botfip-LLM with RWKV-6 as the Teacher LLM.}
    \label{fig:funcimg_OTS_metric}
\end{figure}

Furthermore, we observe the \(Acc_r\), \(S_{RL}\), and \(\tilde{S}_{RL}\) metrics calculated on a validation set of 4000 samples, generated by a symbolic operation tree random generation system, to reflect the training effectiveness and extrapolation ability of the fine-tuned Botfip and Botfip-LLM models (with ChatGLM-2 and RWKV-6 as Teacher LLMs, respectively). Figure \ref{fig:funcimg_OTS_metric} shows the trend of these metrics with varying node counts in the symbolic operation tree for the different models. According to the validation results, the Botfip-LLM model with ChatGLM-2 as the Teacher LLM performed well, slightly surpassing the original Botfip model and the Botfip-LLM model with RWKV-6 as the Teacher in all metrics. Specifically, for the \(Acc_r\) metric, the Botfip-LLM (ChatGLM-2) maintained a higher accuracy throughout the validation process. Although the advantage was not significant, it indicates that the OTS generated by Botfip-LLM (ChatGLM-2) is more effective compared to the original Botfip model and Botfip-LLM (RWKV-6). In terms of \(S_{RL}\) and \(\tilde{S}_{RL}\) metrics, the Botfip-LLM (ChatGLM-2) performed better, showing that the quality of OTS generated by this model is superior to that of the original model and Botfip-LLM (RWKV-6). These results not only confirm the potential of ChatGLM-2 in enhancing model performance but also provide crucial guidance for future model development and experimental design. The outstanding performance of the Botfip-LLM model with ChatGLM-2 as the Teacher model also highlights the importance of selecting an appropriate pre-training model for optimizing training outcomes.

\begin{figure}[!t]
    \centering
    \includegraphics[width=1.0\linewidth]{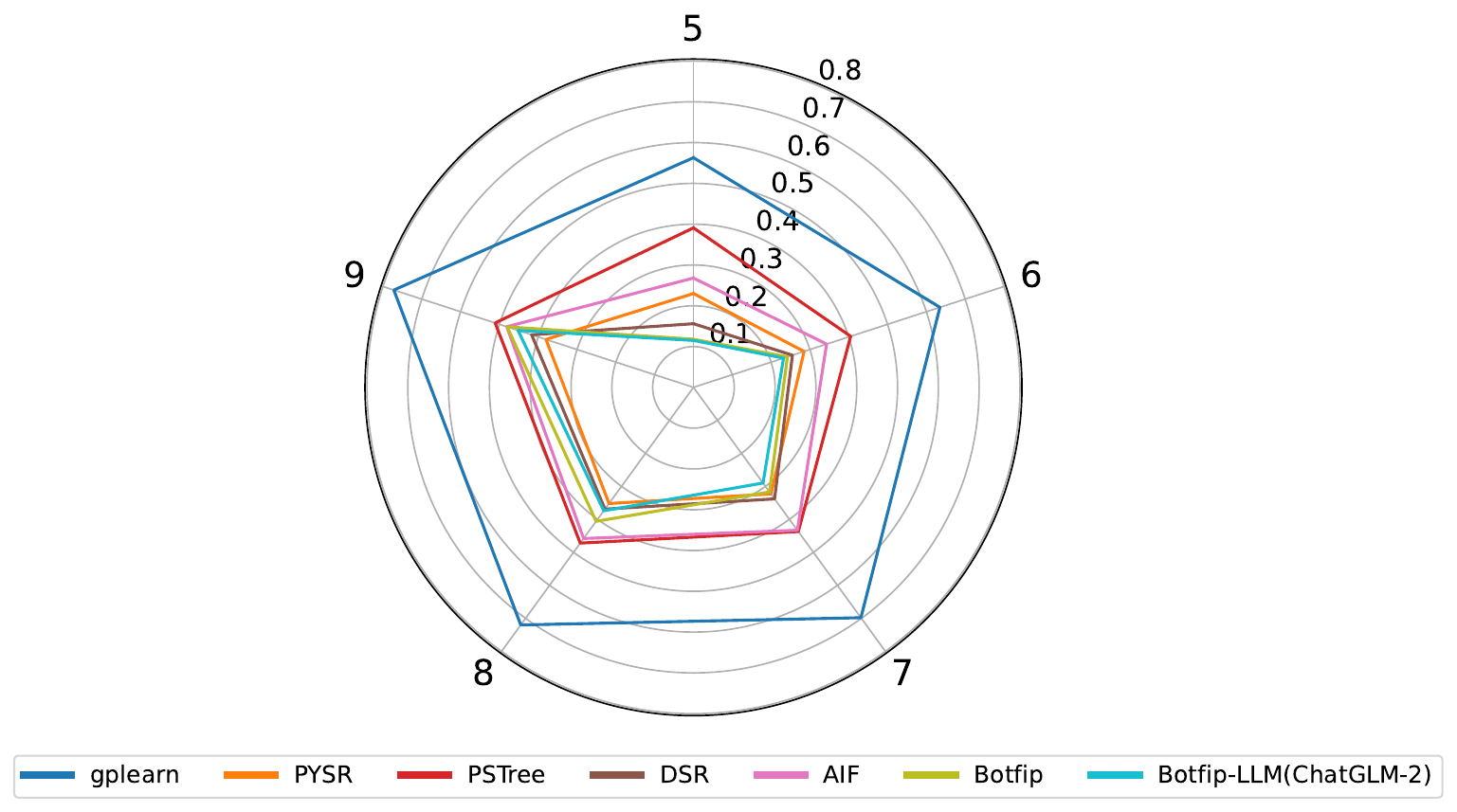}
    \caption{Comparison of MSE results on test sets generated by symbolic operation tree systems with different node counts for various SR models, the original Botfip model, and Botfip-LLM (with ChatGLM-2 as the Teacher LLM).}
    \label{fig:funcimg_OTS_radar}
\end{figure}
Finally, we generated 100 sets of symbolic operation tree skeletons with corresponding constant vectors based on different node counts. The generated function images were rearranged and used as discrete point sets required for symbolic regression methods. These sets were used in the regression process for gplearn, PYSR \cite{cranmer2023interpretable}, PSTree \cite{zhang2022ps}, DSR \cite{landajuela2022unified}, and AI-Feynman (AIF) \cite{udrescu2020ai}. We compared Botfip-LLM and the original Botfip model with these SR frameworks. It is important to note that Botfip-LLM and the original Botfip model are multimodal scientific computing frameworks that rely on function images for feature extraction rather than discrete point sets. Therefore, they cannot be tested using conventional SR datasets. However, the symbolic operation tree system's random generation dataset proposed in \cite{chen2024boOTStrapping} has been open-sourced, and the random generation process is sufficiently reliable. Table \ref{tab: Hyper-parameters of the Dataset and Experiment} reflects part of the dataset, demonstrating its objectivity. Thus, we only consider randomly generated symbolic operation tree data for comparative testing, rather than using conventional SR datasets. The test result information is displayed in the radar chart \ref{fig:funcimg_OTS_radar}. It can be seen that Botfip-LLM shows improved extrapolation ability compared to the original Botfip model. However, as the number of symbolic operation tree nodes increases, the complexity of the OTS also increases. Since Botfip-LLM does not show significant improvements in error correction capabilities compared to the original Botfip model, its extrapolation ability is still inferior to methods like PYSR and DSR, which iteratively search for optimal expressions. Nevertheless, introducing LLM for knowledge distillation has indeed enhanced the model's understanding of symbolic information, further improving recognition and regression performance.

\subsubsection{Formula String-OTS Modeling Task Performance}
In this section, we continue using the same fine-tuning dataset to demonstrate the Symbolic Formula String to OTS modeling fine-tuning task. In this downstream task, the LLM is no longer a Teacher Model but becomes one of the components for feature extraction of multimodal data related to the symbolic operation tree. Similar to the Funcimg-OTS Modeling Task, the LLM processes the Formula String of the symbolic operation tree through a tokenizer to extract symbolic formula features, which are then embedded into the Botfip-LLM main model's OTS Decoder through the corresponding LLM Embedder. By integrating the symbolic formula feature information via Causal Attention, the OTS Decoder can predict the corresponding OTS.

\begin{figure}[!t]
    \centering
    \includegraphics[width=1.0\linewidth]{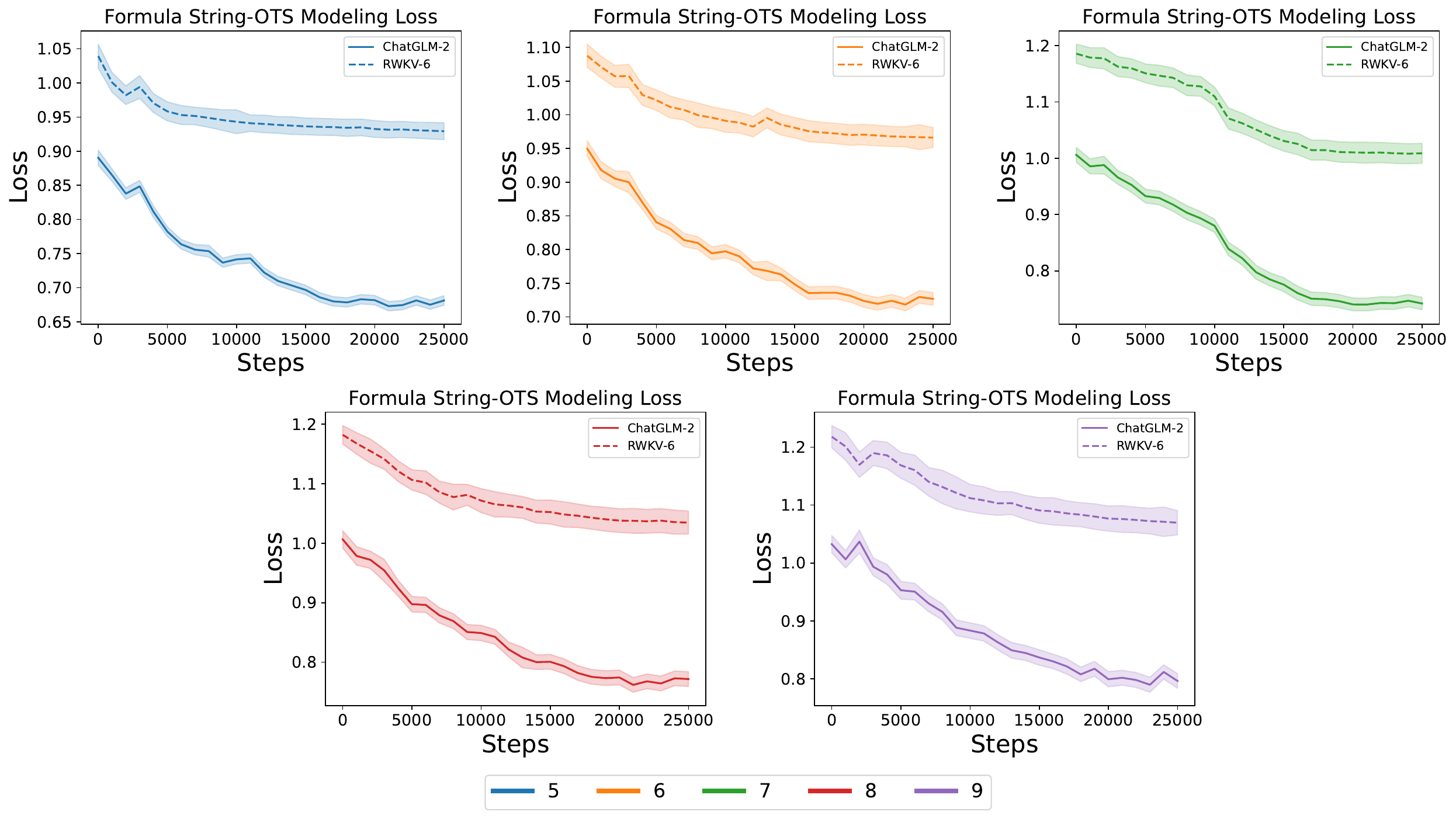}
    \caption{Loss variation during the training process of the Formula String-OTS Modeling fine-tuning task for Botfip-LLM (ChatGLM-2) and Botfip-LLM (RWKV-6). The legend below indicates different colors representing the number of nodes in the symbolic operation trees for the datasets used in the respective fine-tuning experiments.
}
    \label{fig:str_llm_modeling}
\end{figure}

Figure \ref{fig:str_llm_modeling} illustrates the training loss curves for the pre-trained Botfip-LLM models in the Formula String-OTS Modeling fine-tuning task, with ChatGLM-2 and RWKV-6 serving as the Teacher LLMs and corresponding Formula String feature extraction networks. Table \ref{tab:Formula String-OTS Modeling Task loss} shows the respective convergence error conditions. It is evident that in the previous pre-training process, ChatGLM-2 successfully transferred its understanding of symbolic information to the Botfip-LLM main model through knowledge distillation, whereas RWKV-6 did not achieve this effect. Consequently, in this fine-tuning task, the Botfip-LLM model under ChatGLM-2 as the Formula String feature extraction network exhibits a faster decline and better convergence accuracy in Formula String-OTS Modeling loss compared to RWKV-6.
\begin{table}[!t]
\centering
\caption{Mean fine-tuning error and its standard deviation at the \num{25000}th step of Botfip-LLM in Formula String-OTS Modeling Task under ChatGLM-2 and RWKV-6.}
\label{tab:Formula String-OTS Modeling Task loss}
\resizebox{\textwidth}{!}{
\begin{tabular}{cccccc}
\toprule[0.8pt]
\textbf{LLM Name} & node \(5\) & node \(6\) & node \(7\) & node \(8\) & node \(9\)\\ \hline
ChatGLM-2 & \(0.681\pm0.007\) & \(0.727\pm0.009\) & \(0.742\pm0.011\) & \(0.772\pm0.012\) & \(0.796\pm0.013\)\\
RWKV-6 & \(0.929\pm0.012\) & \(0.966\pm0.015\) & \(1.009\pm0.018\) & \(1.034\pm0.019\) & \(1.070\pm0.021\)\\
\bottomrule[0.8pt]
\end{tabular}
}
\end{table}
\begin{figure}[!t]
    \centering
    \includegraphics[width=1.0\linewidth]{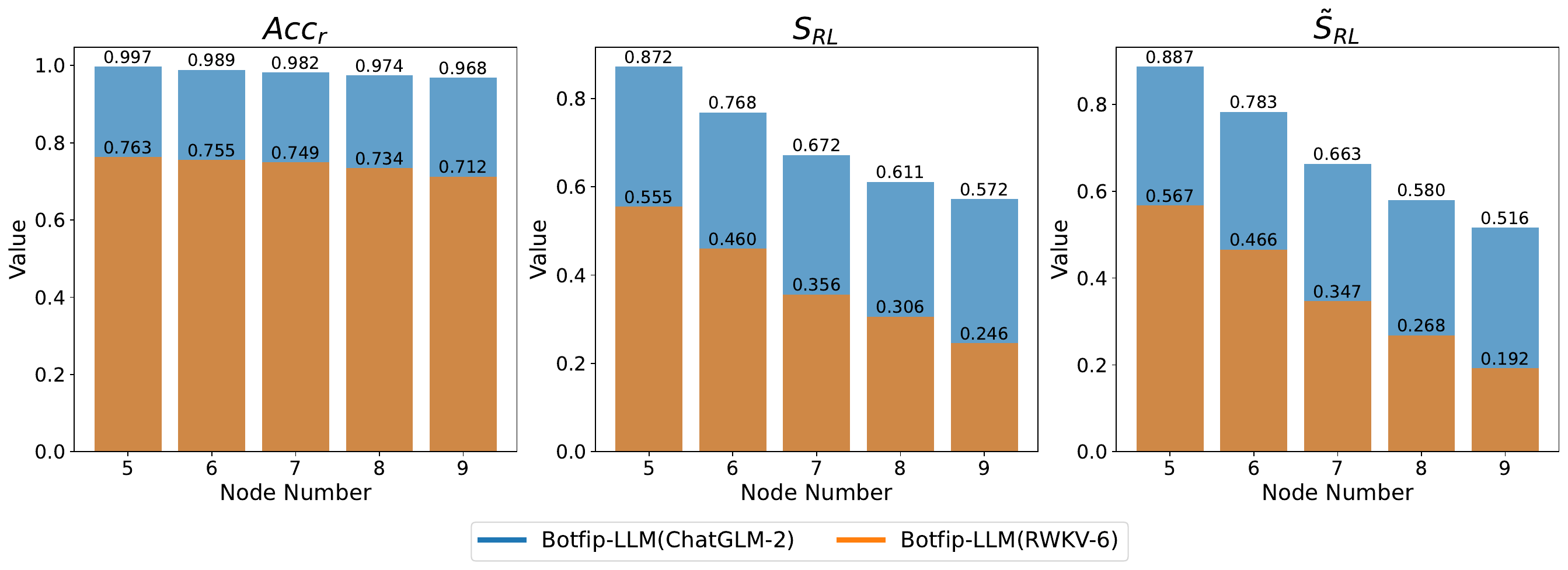}
    \caption{A bar chart visualizing the results on the Formula String-OTS test set after Botfip-LLM (ChatGLM-2) and Botfip-LLM (RWKV-6) underwent fine-tuning for the Formula String-OTS Modeling task.}
    \label{fig:str_llm_bar}
\end{figure}
Now we proceed to test the effectiveness of Botfip-LLM in generating OTS from Formula Strings. We continue to use the three evaluation metrics: \(Acc_{r}\), \(S_{RL}\), and \(\Tilde{S}_{RL}\). Figure \ref{fig:str_llm_bar} presents the test results of Botfip-LLM (ChatGLM-2) and Botfip-LLM (RWKV-6) on 100 randomly generated symbol tree samples after fine-tuning. It can be observed that compared to the Funcimg-OTS fine-tuning task, Botfip-LLM (ChatGLM-2) shows a more significant advantage in the Formula String-OTS fine-tuning task. This primarily stems from the fact that during the pre-training phase, ChatGLM-2 adjusted the parameters of the LLM embedder within Botfip-LLM to a certain extent through fine-tuning, making its dimensionality reduction mapping of the hidden state more suitable for feature extraction of symbolic formula information compared to RWKV-6, thus achieving better performance in this fine-tuning task. From this fine-tuning training, it can also be concluded that the introduction of LLM not only helps the Botfip-LLM framework gain a better understanding of the multimodal information of scientific computing data but also effectively broadens the applicability of the Botfip-LLM framework, thereby further accomplishing downstream tasks related to symbolic formulas.

\section{Conclusion}

This paper proposes the Botfip-LLM multimodal scientific computing framework, which is based on the Botfip framework, which significantly enhances its capability to handle symbolic formula information by introducing LLMs and employing knowledge distillation technology for aligning multimodal data related to symbolic operation trees. Experimental results indicate that the choice of different types of LLM frameworks and their pre-trained weights is crucial for the pre-training and fine-tuning performance of Botfip-LLM. Notably, with the support of ChatGLM-2, Botfip-LLM demonstrates significant improvements in processing and understanding symbolic formula information compared to other Transformer frameworks such as LLaMA-2 and non-Transformer frameworks like RWKV-6 and Mamba. Moreover, by integrating LLMs, the Botfip-LLM framework further extends its range of fine-tuning tasks. Botfip-LLM can not only generate symbolic expressions from OTS but also reverse-engineer OTS sequences and constant vectors from symbolic expressions, significantly enhancing its technical performance and applicability while expanding its use cases in scientific computing tasks. Although Botfip-LLM's extrapolation ability has improved compared to the original Botfip framework, it still exhibits a decline with an increasing number of symbolic operation tree nodes. Future research should focus on further enhancements to improve the model's overall performance and applicability. Botfip-LLM holds promise for helping researchers, engineers, and students better understand the essence and core of scientific computing problems through the alignment and information extraction of function images and symbolic formulas.

\bibliographystyle{unsrt}
\bibliography{refs.bib}

\begin{thebibliography}{10}

\bibitem{conti2024artificial}
Silvia Conti.
\newblock Artificial intelligence for weather forecasting.
\newblock {\em Nature Reviews Electrical Engineering}, 1(1):8--8, 2024.

\bibitem{valanarasu2022transweather}
Jeya Maria~Jose Valanarasu, Rajeev Yasarla, and Vishal~M Patel.
\newblock Transweather: Transformer-based restoration of images degraded by adverse weather conditions.
\newblock In {\em Proceedings of the IEEE/CVF Conference on Computer Vision and Pattern Recognition}, pages 2353--2363, 2022.

\bibitem{ji2024spatio}
Junzhong Ji, Jing He, Minglong Lei, Muhua Wang, and Wei Tang.
\newblock Spatio-temporal transformer network for weather forecasting.
\newblock {\em IEEE Transactions on Big Data}, 2024.

\bibitem{cai2021physics}
Shengze Cai, Zhiping Mao, Zhicheng Wang, Minglang Yin, and George~Em Karniadakis.
\newblock Physics-informed neural networks (pinns) for fluid mechanics: A review.
\newblock {\em Acta Mechanica Sinica}, 37(12):1727--1738, 2021.

\bibitem{raissi2019physics}
Maziar Raissi, Paris Perdikaris, and George~E Karniadakis.
\newblock Physics-informed neural networks: A deep learning framework for solving forward and inverse problems involving nonlinear partial differential equations.
\newblock {\em Journal of Computational physics}, 378:686--707, 2019.

\bibitem{li2020fourier}
Zongyi Li, Nikola Kovachki, Kamyar Azizzadenesheli, Burigede Liu, Kaushik Bhattacharya, Andrew Stuart, and Anima Anandkumar.
\newblock Fourier neural operator for parametric partial differential equations.
\newblock {\em arXiv preprint arXiv:2010.08895}, 2020.

\bibitem{hao2023gnot}
Zhongkai Hao, Zhengyi Wang, Hang Su, Chengyang Ying, Yinpeng Dong, Songming Liu, Ze~Cheng, Jian Song, and Jun Zhu.
\newblock Gnot: A general neural operator transformer for operator learning.
\newblock In {\em International Conference on Machine Learning}, pages 12556--12569. PMLR, 2023.

\bibitem{khan2021gene}
Anwar Khan and Boreom Lee.
\newblock Gene transformer: Transformers for the gene expression-based classification of lung cancer subtypes.
\newblock {\em arXiv preprint arXiv:2108.11833}, 2021.

\bibitem{zhang2022transformer}
Ting-He Zhang, Md~Musaddaqul Hasib, Yu-Chiao Chiu, Zhi-Feng Han, Yu-Fang Jin, Mario Flores, Yidong Chen, and Yufei Huang.
\newblock Transformer for gene expression modeling (t-gem): An interpretable deep learning model for gene expression-based phenotype predictions.
\newblock {\em Cancers}, 14(19):4763, 2022.

\bibitem{kolluri2022machine}
Sheela Kolluri, Jianchang Lin, Rachael Liu, Yanwei Zhang, and Wenwen Zhang.
\newblock Machine learning and artificial intelligence in pharmaceutical research and development: a review.
\newblock {\em The AAPS journal}, 24:1--10, 2022.

\bibitem{chen2023artificial}
Wei Chen, Xuesong Liu, Sanyin Zhang, and Shilin Chen.
\newblock Artificial intelligence for drug discovery: Resources, methods, and applications.
\newblock {\em Molecular Therapy-Nucleic Acids}, 31:691--702, 2023.

\bibitem{fu2022spectratr}
Pengyou Fu, Yue Wen, Yuke Zhang, Lingqiao Li, Yanchun Feng, Lihui Yin, and Huihua Yang.
\newblock Spectratr: A novel deep learning model for qualitative analysis of drug spectroscopy based on transformer structure.
\newblock {\em Journal of Innovative Optical Health Sciences}, 15(03):2250021, 2022.

\bibitem{valipour2021symbolicgpt}
Mojtaba Valipour, Bowen You, Maysum Panju, and Ali Ghodsi.
\newblock Symbolicgpt: A generative transformer model for symbolic regression.
\newblock {\em arXiv preprint arXiv:2106.14131}, 2021.

\bibitem{vastl2024symformer}
Martin Vastl, Jon{\'a}{\v{s}} Kulh{\'a}nek, Ji{\v{r}}{\'\i} Kubal{\'\i}k, Erik Derner, and Robert Babu{\v{s}}ka.
\newblock Symformer: End-to-end symbolic regression using transformer-based architecture.
\newblock {\em IEEE Access}, 2024.

\bibitem{kamienny2022end}
Pierre-Alexandre Kamienny, St{\'e}phane d'Ascoli, Guillaume Lample, and Fran{\c{c}}ois Charton.
\newblock End-to-end symbolic regression with transformers.
\newblock {\em Advances in Neural Information Processing Systems}, 35:10269--10281, 2022.

\bibitem{stevens2020ai}
Rick Stevens, Valerie Taylor, Jeff Nichols, Arthur~Barney Maccabe, Katherine Yelick, and David Brown.
\newblock Ai for science: Report on the department of energy (doe) town halls on artificial intelligence (ai) for science.
\newblock Technical report, Argonne National Lab.(ANL), Argonne, IL (United States), 2020.

\bibitem{radford2021learning}
Alec Radford, Jong~Wook Kim, Chris Hallacy, Aditya Ramesh, Gabriel Goh, Sandhini Agarwal, Girish Sastry, Amanda Askell, Pamela Mishkin, Jack Clark, et~al.
\newblock Learning transferable visual models from natural language supervision.
\newblock In {\em International conference on machine learning}, pages 8748--8763. PMLR, 2021.

\bibitem{li2022blip}
Junnan Li, Dongxu Li, Caiming Xiong, and Steven Hoi.
\newblock Blip: Bootstrapping language-image pre-training for unified vision-language understanding and generation.
\newblock In {\em International conference on machine learning}, pages 12888--12900. PMLR, 2022.

\bibitem{jia2021scaling}
Chao Jia, Yinfei Yang, Ye~Xia, Yi-Ting Chen, Zarana Parekh, Hieu Pham, Quoc Le, Yun-Hsuan Sung, Zhen Li, and Tom Duerig.
\newblock Scaling up visual and vision-language representation learning with noisy text supervision.
\newblock In {\em International conference on machine learning}, pages 4904--4916. PMLR, 2021.

\bibitem{li2021align}
Junnan Li, Ramprasaath Selvaraju, Akhilesh Gotmare, Shafiq Joty, Caiming Xiong, and Steven Chu~Hong Hoi.
\newblock Align before fuse: Vision and language representation learning with momentum distillation.
\newblock {\em Advances in neural information processing systems}, 34:9694--9705, 2021.

\bibitem{chen2024boOTStrapping}
Tianhao Chen, Pengbo Xu, and Haibiao Zheng.
\newblock Bootstrapping ots-funcimg pre-training model (botfip)--a comprehensive symbolic regression framework.
\newblock {\em arXiv preprint arXiv:2401.09748}, 2024.

\bibitem{devlin2018bert}
Jacob Devlin, Ming-Wei Chang, Kenton Lee, and Kristina Toutanova.
\newblock Bert: Pre-training of deep bidirectional transformers for language understanding.
\newblock {\em arXiv preprint arXiv:1810.04805}, 2018.

\bibitem{dosovitskiy2020image}
Alexey Dosovitskiy, Lucas Beyer, Alexander Kolesnikov, Dirk Weissenborn, Xiaohua Zhai, Thomas Unterthiner, Mostafa Dehghani, Matthias Minderer, Georg Heigold, Sylvain Gelly, et~al.
\newblock An image is worth 16x16 words: Transformers for image recognition at scale.
\newblock {\em arXiv preprint arXiv:2010.11929}, 2020.

\bibitem{touvron2023llama}
Hugo Touvron, Thibaut Lavril, Gautier Izacard, Xavier Martinet, Marie-Anne Lachaux, Timoth{\'e}e Lacroix, Baptiste Rozi{\`e}re, Naman Goyal, Eric Hambro, Faisal Azhar, et~al.
\newblock Llama: Open and efficient foundation language models.
\newblock {\em arXiv preprint arXiv:2302.13971}, 2023.

\bibitem{team2024gemma}
Gemma Team, Thomas Mesnard, Cassidy Hardin, Robert Dadashi, Surya Bhupatiraju, Shreya Pathak, Laurent Sifre, Morgane Rivi{\`e}re, Mihir~Sanjay Kale, Juliette Love, et~al.
\newblock Gemma: Open models based on gemini research and technology.
\newblock {\em arXiv preprint arXiv:2403.08295}, 2024.

\bibitem{li2023blip}
Junnan Li, Dongxu Li, Silvio Savarese, and Steven Hoi.
\newblock Blip-2: Bootstrapping language-image pre-training with frozen image encoders and large language models.
\newblock In {\em International conference on machine learning}, pages 19730--19742. PMLR, 2023.

\bibitem{yu2022coca}
Jiahui Yu, Zirui Wang, Vijay Vasudevan, Legg Yeung, Mojtaba Seyedhosseini, and Yonghui Wu.
\newblock Coca: Contrastive captioners are image-text foundation models.
\newblock {\em arXiv preprint arXiv:2205.01917}, 2022.

\bibitem{du2021glm}
Zhengxiao Du, Yujie Qian, Xiao Liu, Ming Ding, Jiezhong Qiu, Zhilin Yang, and Jie Tang.
\newblock Glm: General language model pretraining with autoregressive blank infilling.
\newblock {\em arXiv preprint arXiv:2103.10360}, 2021.

\bibitem{li2023textbooks}
Yuanzhi Li, S{\'e}bastien Bubeck, Ronen Eldan, Allie Del~Giorno, Suriya Gunasekar, and Yin~Tat Lee.
\newblock Textbooks are all you need ii: phi-1.5 technical report.
\newblock {\em arXiv preprint arXiv:2309.05463}, 2023.

\bibitem{peng2023rwkv}
Bo~Peng, Eric Alcaide, Quentin Anthony, Alon Albalak, Samuel Arcadinho, Huanqi Cao, Xin Cheng, Michael Chung, Matteo Grella, Kranthi~Kiran GV, et~al.
\newblock Rwkv: Reinventing rnns for the transformer era.
\newblock {\em arXiv preprint arXiv:2305.13048}, 2023.

\bibitem{gu2023mamba}
Albert Gu and Tri Dao.
\newblock Mamba: Linear-time sequence modeling with selective state spaces.
\newblock {\em arXiv preprint arXiv:2312.00752}, 2023.

\bibitem{patro2024mamba}
Badri~Narayana Patro and Vijay~Srinivas Agneeswaran.
\newblock Mamba-360: Survey of state space models as transformer alternative for long sequence modelling: Methods, applications, and challenges.
\newblock {\em arXiv preprint arXiv:2404.16112}, 2024.

\bibitem{hu2021lora}
Edward~J Hu, Yelong Shen, Phillip Wallis, Zeyuan Allen-Zhu, Yuanzhi Li, Shean Wang, Lu~Wang, and Weizhu Chen.
\newblock Lora: Low-rank adaptation of large language models.
\newblock {\em arXiv preprint arXiv:2106.09685}, 2021.

\bibitem{xu2023wizardlm}
Can Xu, Qingfeng Sun, Kai Zheng, Xiubo Geng, Pu~Zhao, Jiazhan Feng, Chongyang Tao, and Daxin Jiang.
\newblock Wizardlm: Empowering large language models to follow complex instructions.
\newblock {\em arXiv preprint arXiv:2304.12244}, 2023.

\bibitem{wang2022self}
Yizhong Wang, Yeganeh Kordi, Swaroop Mishra, Alisa Liu, Noah~A Smith, Daniel Khashabi, and Hannaneh Hajishirzi.
\newblock Self-instruct: Aligning language models with self-generated instructions.
\newblock {\em arXiv preprint arXiv:2212.10560}, 2022.

\bibitem{sanh2019distilbert}
Victor Sanh, Lysandre Debut, Julien Chaumond, and Thomas Wolf.
\newblock Distilbert, a distilled version of bert: smaller, faster, cheaper and lighter.
\newblock {\em arXiv preprint arXiv:1910.01108}, 2019.

\bibitem{gu2023minillm}
Yuxian Gu, Li~Dong, Furu Wei, and Minlie Huang.
\newblock Minillm: Knowledge distillation of large language models.
\newblock In {\em The Twelfth International Conference on Learning Representations}, 2023.

\bibitem{liang2020mixkd}
Kevin~J Liang, Weituo Hao, Dinghan Shen, Yufan Zhou, Weizhu Chen, Changyou Chen, and Lawrence Carin.
\newblock Mixkd: Towards efficient distillation of large-scale language models.
\newblock {\em arXiv preprint arXiv:2011.00593}, 2020.

\bibitem{timiryasov2023baby}
Inar Timiryasov and Jean-Loup Tastet.
\newblock Baby llama: knowledge distillation from an ensemble of teachers trained on a small dataset with no performance penalty.
\newblock {\em arXiv preprint arXiv:2308.02019}, 2023.

\bibitem{cui2023ultrafeedback}
Ganqu Cui, Lifan Yuan, Ning Ding, Guanming Yao, Wei Zhu, Yuan Ni, Guotong Xie, Zhiyuan Liu, and Maosong Sun.
\newblock Ultrafeedback: Boosting language models with high-quality feedback.
\newblock {\em arXiv preprint arXiv:2310.01377}, 2023.

\bibitem{kwon2023reward}
Minae Kwon, Sang~Michael Xie, Kalesha Bullard, and Dorsa Sadigh.
\newblock Reward design with language models.
\newblock {\em arXiv preprint arXiv:2303.00001}, 2023.

\bibitem{tunstall2023zephyr}
Lewis Tunstall, Edward Beeching, Nathan Lambert, Nazneen Rajani, Kashif Rasul, Younes Belkada, Shengyi Huang, Leandro von Werra, Cl{\'e}mentine Fourrier, Nathan Habib, et~al.
\newblock Zephyr: Direct distillation of lm alignment.
\newblock {\em arXiv preprint arXiv:2310.16944}, 2023.

\bibitem{hong2023cyclealign}
Jixiang Hong, Quan Tu, Changyu Chen, Xing Gao, Ji~Zhang, and Rui Yan.
\newblock Cyclealign: Iterative distillation from black-box llm to white-box models for better human alignment.
\newblock {\em arXiv preprint arXiv:2310.16271}, 2023.

\bibitem{xu2024survey}
Xiaohan Xu, Ming Li, Chongyang Tao, Tao Shen, Reynold Cheng, Jinyang Li, Can Xu, Dacheng Tao, and Tianyi Zhou.
\newblock A survey on knowledge distillation of large language models.
\newblock {\em arXiv preprint arXiv:2402.13116}, 2024.

\bibitem{petersen2019deep}
Brenden~K Petersen, Mikel Landajuela, T~Nathan Mundhenk, Claudio~P Santiago, Soo~K Kim, and Joanne~T Kim.
\newblock Deep symbolic regression: Recovering mathematical expressions from data via risk-seeking policy gradients.
\newblock {\em arXiv preprint arXiv:1912.04871}, 2019.

\bibitem{zhang2021rl}
Hengzhe Zhang and Aimin Zhou.
\newblock Rl-gep: symbolic regression via gene expression programming and reinforcement learning.
\newblock In {\em 2021 International Joint Conference on Neural Networks (IJCNN)}, pages 1--8. IEEE, 2021.

\bibitem{crochepierre2022interactive}
Laure Crochepierre, Lydia Boudjeloud-Assala, and Vincent Barbesant.
\newblock Interactive reinforcement learning for symbolic regression from multi-format human-preference feedbacks.
\newblock In {\em IJCAI}, pages 5900--5903, 2022.

\bibitem{shojaee2024transformer}
Parshin Shojaee, Kazem Meidani, Amir Barati~Farimani, and Chandan Reddy.
\newblock Transformer-based planning for symbolic regression.
\newblock {\em Advances in Neural Information Processing Systems}, 36, 2024.

\bibitem{biggio2020seq2seq}
Luca Biggio, Tommaso Bendinelli, Aurelien Lucchi, and Giambattista Parascandolo.
\newblock A seq2seq approach to symbolic regression.
\newblock In {\em Learning Meets Combinatorial Algorithms at NeurIPS2020}, 2020.

\bibitem{arechiga2021accelerating}
Nikos Ar{\'e}chiga, Francine Chen, Yan-Ying Chen, Yanxia Zhang, Rumen Iliev, Heishiro Toyoda, and Kent Lyons.
\newblock Accelerating understanding of scientific experiments with end to end symbolic regression.
\newblock {\em arXiv preprint arXiv:2112.04023}, 2021.

\bibitem{chen2020improved}
Xinlei Chen, Haoqi Fan, Ross Girshick, and Kaiming He.
\newblock Improved baselines with momentum contrastive learning.
\newblock {\em arXiv preprint arXiv:2003.04297}, 2020.

\bibitem{yujian2007normalized}
Li~Yujian and Liu Bo.
\newblock A normalized levenshtein distance metric.
\newblock {\em IEEE transactions on pattern analysis and machine intelligence}, 29(6):1091--1095, 2007.

\bibitem{zeng2022glm}
Aohan Zeng, Xiao Liu, Zhengxiao Du, Zihan Wang, Hanyu Lai, Ming Ding, Zhuoyi Yang, Yifan Xu, Wendi Zheng, Xiao Xia, et~al.
\newblock Glm-130b: An open bilingual pre-trained model.
\newblock {\em arXiv preprint arXiv:2210.02414}, 2022.

\bibitem{jiang2023mistral}
Albert~Q Jiang, Alexandre Sablayrolles, Arthur Mensch, Chris Bamford, Devendra~Singh Chaplot, Diego de~las Casas, Florian Bressand, Gianna Lengyel, Guillaume Lample, Lucile Saulnier, et~al.
\newblock Mistral 7b.
\newblock {\em arXiv preprint arXiv:2310.06825}, 2023.

\bibitem{bai2023qwen}
Jinze Bai, Shuai Bai, Yunfei Chu, Zeyu Cui, Kai Dang, Xiaodong Deng, Yang Fan, Wenbin Ge, Yu~Han, Fei Huang, et~al.
\newblock Qwen technical report.
\newblock {\em arXiv preprint arXiv:2309.16609}, 2023.

\bibitem{cranmer2023interpretable}
Miles Cranmer.
\newblock Interpretable machine learning for science with pysr and symbolicregression.jl.
\newblock {\em arXiv preprint arXiv:2305.01582}, 2023.

\bibitem{zhang2022ps}
Hengzhe Zhang, Aimin Zhou, Hong Qian, and Hu~Zhang.
\newblock Ps-tree: A piecewise symbolic regression tree.
\newblock {\em Swarm and Evolutionary Computation}, 71:101061, 2022.

\bibitem{landajuela2022unified}
Mikel Landajuela, Chak~Shing Lee, Jiachen Yang, Ruben Glatt, Claudio~P Santiago, Ignacio Aravena, Terrell Mundhenk, Garrett Mulcahy, and Brenden~K Petersen.
\newblock A unified framework for deep symbolic regression.
\newblock {\em Advances in Neural Information Processing Systems}, 35:33985--33998, 2022.

\bibitem{udrescu2020ai}
Silviu-Marian Udrescu and Max Tegmark.
\newblock Ai feynman: A physics-inspired method for symbolic regression.
\newblock {\em Science Advances}, 6(16):eaay2631, 2020.

\end{thebibliography}
\newpage
\appendix
\section{Hyper-parameters Involved in the Dataset and Experiment}\label{apd:Hyper-parameters Involved in the Dataset and Experiment}

\begin{table}[!th]
\centering
\caption{Hyper-parameters of the dataset, model, and training process}
\label{tab: Hyper-parameters of the Dataset and Experiment}
\begin{tabular}{ll}
\toprule[0.8pt]
\textbf{Parameter} & \textbf{Description} \\ \hline
OTS Encoder/Decoder Framework & 10-layer BERT \\ 
OTS Encoder/Decoder Hidden Size & 768 \\ 
Funcimg Encoder & 10-layer ViT \\ 
Funcimg Encoder Hidden  Size & 768 \\ 
Node Ranges & \([5,15]\)\\
Number of OTS (Pre-training) & 11028 \\ 
Number of Funcimg-OTS Pairs (Pre-training) & 551400 \\ 
Sampling Range of Operation Tree Constants Array & \([-2,2]\) \\ 
Number of OTS Skeletons (Fine-tuning \& validation) & 2941 \\ 
Number of Funcimg-OTS Pairs (Fine-tuning \& validation) & 88230 \\ 
Number of Funcimg-OTS Pairs used in  validation  & 4000 \\
Learning Rate Adjustment Method & StepLR \\ 
Initial Learning Rate During Warm-up & \(1 \times 10^{-6}\) \\ 
Maximum Initial Learning Rate & \(1 \times 10^{-4}\) \\ 
Function Images Input Noise & 0.001\\
Learning rate decay rate  & \(0.9\) \\ 
Learning rate decay period& \(5\) epochs \\ 
Total Duration of Training & 100 epochs \\ 
GPU device &  Nvidia A6000 GPU \\ 
Number of GPU used in Pre-training for Main Model  & 4  \\ 
Number of GPU for LLM deployment & 1  \\ 
Number of GPU used in Fine-tuning & 1 \\  \bottomrule[0.8pt]
\end{tabular}
\end{table}

\newpage
\section{Appendix: Selected Skeletons Information from the Funcimg-OTS-Formula String Dataset}

\label{sec: Appendix of Display Selected Skeleton Information}

\begin{longtable}{>{\small}c>{\small}l>{\small}l}
\caption{Selected Formula Skeletons and Their OTS from the Funcimg-OTS Dataset}\\
\hline
\multicolumn{1}{c}{\textbf{{Nodes}}} & \multicolumn{1}{c}{\textbf{Formula Skeleton}} & \multicolumn{1}{c}{\textbf{OTS}} \\
\hline
\endfirsthead
\hline
\textbf{\makecell[l]{Nodes}} & \textbf{Formula Skeleton} & \textbf{OTS} \\
\hline
\endhead
\hline \multicolumn{3}{r}{{Continued on next page}} \\

\endfoot
    
    \multicolumn{3}{@{}p{\linewidth}@{}}{
        \fontsize{10}{12}\selectfont
            $^{*}$  The table shows only a subset of our datasets; in fact, our pretraining dataset contains 11,028 skeletons and 551,400 Funcimg-OTS pairs. All constant vectors were sampled from the uniform distribution \( U(-20, 20) \).
      }
    \endlastfoot

5 & \(C_0 \cdot (x_0 + x_1)^{2} + C_1\) & \([15, 0, 1, 0, 19, 18, 0, 0, 0]\) \\ 
5 & \(C_0 \cdot \lvert x_0 x_1 \rvert + C_1\) & \([20, 0, 3, 0, 18, 19, 0, 0, 0]\) \\ 
5 & \(C_0 \cdot \frac{C_2}{x_0 x_1} + C_1\) & \([6, 0, 3, 0, 19, 18, 0, 0, 0]\) \\ 
5 & \(C_0 \cdot (x_0 - x_1)^{2} + C_1\) & \([15, 0, 2, 0, 18, 19, 0, 0, 0]\) \\ 
5 & \(C_0 \cdot \tanh(x_0 + x_1) + C_1\) & \([9, 0, 1, 0, 18, 19, 0, 0, 0]\) \\ 
5 & \(C_0 \cdot \sin(x_0 + x_1) + C_1\) & \([5, 0, 1, 0, 19, 18, 0, 0, 0]\) \\ 
5 & \(C_0 \cdot \frac{x_0^2}{x_1^2} + C_1\) & \([15, 0, 4, 0, 18, 19, 0, 0, 0]\) \\ 
5 & \(C_0 \cdot \tan(x_0 x_1) + C_1\) & \([7, 0, 3, 0, 18, 19, 0, 0, 0]\) \\ 
5 & \(C_0 \cdot \frac{C_2}{\sin(\exp(x_0 x_1))} + C_1\) & \([6, 0, 5, 0, 11, 0, 3, 0, 18, 19, 0, 0, 0]\) \\ 
5 & \(C_0 \cdot \exp(x_1 / x_0) + C_1\) & \([11, 0, 4, 0, 19, 18, 0, 0, 0]\) \\ 
5 & \(C_0 \cdot x_1 \cdot \lvert x_0 \rvert + C_1\) & \([3, 0, 20, 19, 0, 18, 0, 0, 0]\) \\ 
5 & \(C_0 \cdot (C_2 / x_0 + x_1) + C_1\) & \([1, 0, 6, 19, 0, 18, 0, 0, 0]\) \\ 
5 & \(C_0 \cdot (-x_1 + \exp(x_0)) + C_1\) & \([2, 0, 11, 19, 0, 18, 0, 0, 0]\) \\ 
5 & \(C_0 \cdot \log(\lvert x_1 \rvert) / x_0 + C_1\) & \([4, 0, 12, 18, 0, 20, 0, 0, 19, 0, 0]\) \\ 
5 & \(C_0 \cdot x_1 \cdot \sqrt{\lvert \sin(x_0) \rvert} + C_1\) & \([3, 0, 16, 19, 0, 5, 0, 0, 18, 0, 0]\) \\ 
5 & \(C_0 \cdot (x_1 + \tanh(x_0)) + C_1\) & \([1, 0, 9, 19, 0, 18, 0, 0, 0]\) \\ 
5 & \(C_0 \cdot \sin(x_0 - x_1) + C_1\) & \([5, 0, 2, 0, 18, 19, 0, 0, 0]\) \\ 
5 & \(C_0 \cdot \lvert x_1 / x_0 \rvert + C_1\) & \([20, 0, 4, 0, 19, 18, 0, 0, 0]\) \\ 
5 & \(C_0 \cdot (x_0 + \sin(x_1)) + C_1\) & \([1, 0, 5, 18, 0, 19, 0, 0, 0]\) \\ 
5 & \(C_0 \cdot \tanh(x_0 x_1) + C_1\) & \([9, 0, 3, 0, 19, 18, 0, 0, 0]\) \\ 
5 & \(C_0 \cdot \sqrt{\lvert x_0 \rvert} + C_1\) & \([16, 0, 3, 0, 18, 18, 0, 0, 0]\) \\ 
5 & \(C_0 \cdot \exp(x_1 / x_0) + C_1\) & \([11, 0, 4, 0, 18, 19, 0, 0, 0]\) \\ 
5 & \(C_0 \cdot \tan(x_0 - x_1) + C_1\) & \([7, 0, 2, 0, 18, 19, 0, 0, 0]\) \\ 
5 & \(C_0 \cdot x_0^2 x_1^2 + C_1\) & \([15, 0, 3, 0, 19, 18, 0, 0, 0]\) \\ 
5 & \(C_0 \cdot \sin(x_0) / x_1 + C_1\) & \([4, 0, 5, 19, 0, 18, 0, 0, 0]\) \\ 
5 & \(C_0 \cdot x_0 \cdot \log(\lvert x_1 \rvert) + C_1\) & \([3, 0, 12, 18, 0, 19, 0, 0, 0]\) \\ 
5 & \(C_0 \cdot \exp(-x_0 + x_1) + C_1\) & \([11, 0, 2, 0, 19, 18, 0, 0, 0]\) \\ 
5 & \(C_0 \cdot (x_0 + \tan(x_1)) + C_1\) & \([1, 0, 7, 18, 0, 19, 0, 0, 0]\) \\ 
5 & \(C_0 \cdot (x_0 + \exp(x_1)) + C_1\) & \([1, 0, 11, 18, 0, 19, 0, 0, 0]\) \\ 
5 & \(C_0 \cdot (x_1 + \sin(x_0)) + C_1\) & \([1, 0, 5, 19, 0, 18, 0, 0, 0]\) \\ 
6 & \(C_0 \cdot (-x_1^2 + \tan(x_0)) + C_1\) & \([2, 0, 7, 15, 0, 18, 0, 19, 0, 0, 0]\) \\ 
6 & \(C_0 \cdot (\exp(x_1) + \tan(x_1)) + C_1\) & \([1, 0, 7, 11, 0, 19, 0, 19, 0, 0, 0]\) \\ 
6 & \(C_0 \cdot (\exp(x_1) + \sin(x_0)) + C_1\) & \([1, 0, 11, 5, 0, 19, 0, 18, 0, 0, 0]\) \\ 
6 & \(C_0 \cdot (-C_2 x_1 - C_3 + \sin(x_0)) + C_1\) & \([2, 0, 5, 13, 0, 18, 0, 19, 0, 0, 0]\) \\ 
6 & \(C_0 \cdot (\exp(x_0) + \tanh(x_1)) + C_1\) & \([1, 0, 11, 9, 0, 18, 0, 19, 0, 0, 0]\) \\ 
6 & \(C_0 \cdot (x_0 - x_1)^{2 C_2} + C_1\) & \([15, 0, 14, 0, 2, 0, 18, 19, 0, 0, 0]\) \\ 
6 & \(C_0 \cdot \lvert C_2 x_0 x_1 + C_3 \rvert + C_1\) & \([20, 0, 13, 0, 3, 0, 19, 18, 0, 0, 0]\) \\ 
6 & \(C_0 \cdot \tanh(x_0 + x_1)^2 + C_1\) & \([15, 0, 9, 0, 1, 0, 18, 19, 0, 0, 0]\) \\ 
6 & \(C_0 \cdot \tanh(x_0 x_1)^2 + C_1\) & \([15, 0, 9, 0, 3, 0, 19, 18, 0, 0, 0]\) \\ 
6 & \(C_0 \cdot \lvert \frac{C_2}{x_0 x_1} \rvert + C_1\) & \([20, 0, 6, 0, 3, 0, 18, 19, 0, 0, 0]\) \\ 
6 & \(C_0 \cdot \tan(\sin(x_0 x_1)) + C_1\) & \([7, 0, 5, 0, 3, 0, 18, 19, 0, 0, 0]\) \\ 
6 & \(C_0 \cdot \exp(\max(0, x_0 + x_1)) + C_1\) & \([11, 0, 21, 0, 1, 0, 18, 19, 0, 0, 0]\) \\ 
6 & \(C_0 \cdot \sin(\sqrt{\lvert x_1^2 \rvert}) + C_1\) & \([5, 0, 16, 0, 3, 0, 19, 19, 0, 0, 0]\) \\ 
6 & \(C_0 \cdot \sqrt{\lvert \tan(x_0 x_1) \rvert} + C_1\) & \([16, 0, 7, 0, 3, 0, 18, 19, 0, 0, 0]\) \\ 
6 & \(C_0 \cdot \frac{C_2}{(x_0 + x_1)^2} + C_1\) & \([6, 0, 15, 0, 1, 0, 19, 18, 0, 0, 0]\) \\ 
6 & \(C_0 \cdot \frac{C_2}{C_3 (x_0 - x_1) + C_4} + C_1\) & \([6, 0, 13, 0, 2, 0, 18, 19, 0, 0, 0]\) \\ 
6 & \(C_0 \cdot \tan(C_2 x_0 x_1 + C_3) + C_1\) & \([7, 0, 13, 0, 3, 0, 18, 19, 0, 0, 0]\) \\ 
6 & \(C_0 \cdot (x_1^2 - \exp(x_0)) + C_1\) & \([2, 0, 15, 11, 0, 19, 0, 18, 0, 0, 0]\) \\ 
6 & \(C_0 \cdot x_0^2 \exp(x_0) + C_1\) & \([3, 0, 15, 11, 0, 18, 0, 18, 0, 0, 0]\) \\ 
6 & \(C_0 \cdot (x_0^2 + \sin(x_0)) + C_1\) & \([1, 0, 5, 15, 0, 18, 0, 18, 0, 0, 0]\) \\ 
6 & \(C_0 \cdot \sin(x_1) \tanh(x_1) + C_1\) & \([3, 0, 9, 5, 0, 19, 0, 19, 0, 0, 0]\) \\ 
6 & \(C_0 \cdot (C_2 / x_0 + \sin(x_1)) + C_1\) & \([1, 0, 5, 6, 0, 19, 0, 18, 0, 0, 0]\) \\ 
6 & \(C_1 \cdot (-C_0 + x_0 - x_1) + C_2\) & \([2, 0, 2, 10, 0, 18, 19, 0, 0, 0, 0]\) \\ 
6 & \(C_0 \cdot (-x_0 + 2 x_1) + C_1\) & \([1, 0, 2, 19, 0, 19, 18, 0, 0, 0, 0]\) \\ 
6 & \(C_0 \cdot (-2 x_0 + x_1) + C_1\) & \([2, 0, 2, 18, 0, 19, 18, 0, 0, 0, 0]\) \\ 
6 & \(C_0 \cdot (-x_0 + x_1^{1/3})^2 + C_1\) & \([15, 0, 2, 0, 17, 18, 0, 19, 0, 0, 0]\) \\
7 & \( C_0 \cdot \log(\left|x_1 + \exp(x_0)\right|) + C_1 \) & \([12, 0, 20, 0, 1, 0, 11, 19, 0, 18, 0, 0, 0]\) \\
7 & \( C_0 \cdot \tan(\sqrt{\left|x_0 - \exp(x_1)\right|}) + C_1 \) & \([7, 0, 16, 0, 2, 0, 11, 18, 0, 19, 0, 0, 0]\) \\
7 & \( C_0 \cdot \left|C_2^2 \cdot \frac{x_0^2}{x_1^2}\right| + C_1 \) & \([20, 0, 15, 0, 3, 0, 6, 18, 0, 19, 0, 0, 0]\) \\
7 & \( C_0 \cdot \log\left(\left|\frac{\tanh(x_1)^2}{x_0^2}\right|\right) + C_1 \) & \([12, 0, 15, 0, 4, 0, 9, 18, 0, 19, 0, 0, 0]\) \\
7 & \( C_0 \cdot \tanh(\left|C_2 \cdot x_1 + C_3 + x_0\right|) + C_1 \) & \([9, 0, 20, 0, 1, 0, 13, 18, 0, 19, 0, 0, 0]\) \\
7 & \( C_0 \cdot \frac{C_2^2}{(x_0 + \tan(x_1))^2} + C_1 \) & \([15, 0, 6, 0, 1, 0, 7, 18, 0, 19, 0, 0, 0]\) \\
7 & \( C_0 \cdot C_1 \cdot x_1 \cdot \tanh(x_0) + C_2 \) & \([3, 0, 9, 3, 0, 18, 0, 10, 19, 0, 0, 0, 0]\) \\
7 & \( C_0 \cdot \frac{C_2 \cdot x_1 + C_3}{x_0 - x_1} + C_1 \) & \([4, 0, 13, 2, 0, 19, 0, 18, 19, 0, 0, 0, 0]\) \\
7 & \( C_1 \cdot (-C_0 + x_0) \cdot \sin(x_1) + C_2 \) & \([3, 0, 5, 2, 0, 19, 0, 18, 10, 0, 0, 0, 0]\) \\
7 & \( C_1 \cdot (-C_0 + x_1 + \tanh(x_0)) + C_2 \) & \([2, 0, 9, 2, 0, 18, 0, 10, 19, 0, 0, 0, 0]\) \\
7 & \( C_0^2 \cdot C_1 \cdot (x_0 + x_1)^2 + C_2 \) & \([15, 0, 3, 0, 1, 10, 0, 19, 18, 0, 0, 0, 0]\) \\
7 & \( C_1 \cdot \sin(C_0 - x_0 + x_1) + C_2 \) & \([5, 0, 2, 0, 1, 18, 0, 10, 19, 0, 0, 0, 0]\) \\
7 & \( C_1 \cdot \tanh(C_0 + x_0 - x_1) + C_2 \) & \([9, 0, 1, 0, 2, 10, 0, 18, 19, 0, 0, 0, 0]\) \\
7 & \( C_1 \cdot \tanh\left(\frac{x_0 + x_1}{C_0}\right) + C_2 \) & \([9, 0, 4, 0, 1, 10, 0, 19, 18, 0, 0, 0, 0]\) \\
7 & \( C_0 \cdot (\exp(x_1) + \sqrt{\left|x_1\right|}) + C_1 \) & \([1, 0, 16, 11, 0, 20, 0, 19, 0, 19, 0, 0, 0]\) \\
7 & \( C_0 \cdot (x_1^2 + \exp(\tan(x_1))) + C_1 \) & \([1, 0, 11, 15, 0, 7, 0, 19, 0, 19, 0, 0, 0]\) \\
7 & \( C_0 \cdot (x_1^2 + \sqrt{\left|C_2 / x_1\right|}) + C_1 \) & \([1, 0, 16, 15, 0, 6, 0, 19, 0, 19, 0, 0, 0]\) \\
7 & \( C_0 \cdot (-\sin(x_1) + \left|x_0\right|^2) + C_1 \) & \([2, 0, 15, 5, 0, 20, 0, 19, 0, 18, 0, 0, 0]\) \\
7 & \( C_0 \cdot (x_0^2 + \sin(x_1)^2) + C_1 \) & \([1, 0, 15, 15, 0, 5, 0, 18, 0, 19, 0, 0, 0]\) \\
7 & \( C_0 \cdot (-\log(\left|x_0\right|) + \left|C_2 \cdot x_1 + C_3\right|) + C_1 \) & \([2, 0, 20, 12, 0, 13, 0, 18, 0, 19, 0, 0, 0]\) \\
7 & \( C_0 \cdot \exp(-x_0) \cdot \sin(\tan(x_0)) + C_1 \) & \([4, 0, 5, 11, 0, 7, 0, 18, 0, 18, 0, 0, 0]\) \\
7 & \( C_0 \cdot (\exp(x_0) + \left|C_2 / x_1\right|) + C_1 \) & \([1, 0, 20, 11, 0, 6, 0, 18, 0, 19, 0, 0, 0]\) \\
7 & \( C_0 \cdot (x_0 \cdot x_1 + \sin(x_0)) + C_1 \) & \([1, 0, 3, 5, 0, 18, 19, 0, 18, 0, 0, 0, 0]\) \\
7 & \( C_0 \cdot \frac{C_2^2}{(-x_0 + \exp(x_1))^2} + C_1 \) & \([15, 0, 6, 0, 2, 0, 11, 18, 0, 19, 0, 0, 0]\) \\
7 & \( C_0 \cdot \frac{C_2}{\sqrt{\left|x_0 + \max(0, x_1)\right|}} + C_1 \) & \([6, 0, 16, 0, 1, 0, 21, 18, 0, 19, 0, 0, 0]\) \\
7 & \( C_0 \cdot \sqrt{\left|\tan\left(\frac{C_2}{x_0} + x_1\right)\right|} + C_1 \) & \([16, 0, 7, 0, 1, 0, 6, 19, 0, 18, 0, 0, 0]\) \\
7 & \( C_0 \cdot (C_2 \cdot (-x_0 + x_1^2) + C_3)^2 + C_1 \) & \([15, 0, 13, 0, 2, 0, 15, 18, 0, 19, 0, 0, 0]\) \\
7 & \( C_0 \cdot \exp(2 \cdot \text{re}(x_0)) \cdot \left|x_1\right|^2 + C_1 \) & \([15, 0, 20, 0, 3, 0, 11, 19, 0, 18, 0, 0, 0]\) \\
7 & \( C_0 \cdot (-x_0 + x_1 + \sqrt{\left|x_1\right|}) + C_1 \) & \([2, 0, 16, 2, 0, 19, 0, 18, 19, 0, 0, 0, 0]\) \\
7 & \( C_0 \cdot (C_2 \cdot x_0 + C_3 + x_0 - x_1) + C_1 \) & \([1, 0, 13, 2, 0, 18, 0, 18, 19, 0, 0, 0, 0]\) \\
7 & \( C_0 \cdot x_0^{\frac{1}{3}} \cdot (x_0 + x_1) + C_1 \) & \([3, 0, 17, 1, 0, 18, 0, 18, 19, 0, 0, 0, 0]\) \\
7 & \( C_1 \cdot (-C_0 + C_3 \cdot x_1 + C_4 - x_0) + C_2 \) & \([2, 0, 13, 1, 0, 19, 0, 10, 18, 0, 0, 0, 0]\) \\
7 & \( C_0 \cdot \left(\frac{x_0}{x_1} + x_1^2\right) + C_1 \) & \([1, 0, 15, 4, 0, 19, 0, 18, 19, 0, 0, 0, 0]\) \\
7 & \( C_0 \cdot C_1 \cdot x_1 \cdot \tanh(x_0) + C_2 \) & \([3, 0, 9, 3, 0, 18, 0, 19, 10, 0, 0, 0, 0]\) \\
7 & \( C_1 \cdot (C_0 + x_0) \cdot \tan(x_1) + C_2 \) & \([3, 0, 7, 1, 0, 19, 0, 10, 18, 0, 0, 0, 0]\) \\
7 & \( C_0 \cdot \frac{C_2 \cdot x_0 + C_3}{x_0 \cdot x_1} + C_1 \) & \([4, 0, 13, 3, 0, 18, 0, 19, 18, 0, 0, 0, 0]\) \\
7 & \( C_0 \cdot (-x_0 + x_1) \cdot \tanh(x_0) + C_1 \) & \([3, 0, 9, 2, 0, 18, 0, 19, 18, 0, 0, 0, 0]\) \\
8 & \( C_0 \cdot \exp(x_1) \cdot \tan(x_1) / x_0 + C_1 \) & \([3, 0, 4, 11, 0, 7, 18, 0, 19, 0, 19, 0, 0, 0, 0]\) \\
8 & \( C_0 \cdot x_1 \cdot \left(\frac{C_3}{x_0} + x_0\right) / C_2 + C_1 \) & \([4, 0, 1, 6, 0, 6, 18, 0, 19, 0, 18, 0, 0, 0, 0]\) \\
8 & \( C_0 \cdot (x_0 + \sin(x_1) - \tanh(x_0)) + C_1 \) & \([2, 0, 1, 9, 0, 5, 18, 0, 18, 0, 19, 0, 0, 0, 0]\) \\
8 & \( C_1 \cdot (-C_0 + \tan(x_1) + \max(0, x_0)) + C_2 \) & \([1, 0, 2, 21, 0, 7, 10, 0, 18, 0, 19, 0, 0, 0, 0]\) \\
8 & \( C_0 \cdot \left(\frac{C_2}{x_0} + x_0^2 + x_1\right) + C_1 \) & \([1, 0, 1, 6, 0, 15, 19, 0, 18, 0, 18, 0, 0, 0, 0]\) \\
8 & \( C_0 \cdot (x_0 \cdot x_1 - x_1^2)^{C_2} + C_1 \) & \([14, 0, 2, 0, 3, 15, 0, 18, 19, 0, 19, 0, 0, 0, 0]\) \\
8 & \( C_1 \cdot \tan\left(\frac{C_0}{x_1} - \sin(x_0)\right) + C_2 \) & \([7, 0, 2, 0, 4, 5, 0, 10, 19, 0, 18, 0, 0, 0, 0]\) \\
8 & \( C_0 \cdot \exp(x_0 \cdot x_1 - x_1^2) + C_1 \) & \([11, 0, 2, 0, 3, 15, 0, 18, 19, 0, 19, 0, 0, 0, 0]\) \\
8 & \( C_1 \cdot \left|\frac{C_0 \cdot x_1}{\max(0, x_0)}\right| + C_2 \) & \([20, 0, 4, 0, 3, 21, 0, 10, 19, 0, 18, 0, 0, 0, 0]\) \\
8 & \( \sqrt{2} \cdot C_0 \cdot \sqrt{\left|x_0 \cdot (C_2 \cdot x_1 + C_3)\right|} + C_1 \) & \([16, 0, 3, 0, 1, 13, 0, 18, 18, 0, 19, 0, 0, 0, 0]\) \\
8 & \( C_0 \cdot \left(C_3 \cdot \sqrt{\left|x_1\right|} / x_1\right)^{2 C_2} + C_1 \) & \([15, 0, 14, 0, 3, 0, 6, 16, 0, 19, 0, 19, 0, 0, 0]\) \\
8 & \( C_0 \cdot \left|C_2 \cdot x_1 + C_3 + \tanh(x_1)\right| + C_1 \) & \([21, 0, 20, 0, 1, 0, 13, 9, 0, 19, 0, 19, 0, 0, 0]\) \\
8 & \( C_0 \cdot \left|(x_1^{C_2} - \exp(x_0))^2\right| + C_1 \) & \([20, 0, 15, 0, 2, 0, 14, 11, 0, 19, 0, 18, 0, 0, 0]\) \\
8 & \( C_0 \cdot \left|\exp(x_1) - \tanh(x_1)\right|^2 + C_1 \) & \([15, 0, 20, 0, 2, 0, 9, 11, 0, 19, 0, 19, 0, 0, 0]\) \\
8 & \( C_0 \cdot \tan(\tan(x_1) + \left|x_1\right|)^2 + C_1 \) & \([15, 0, 7, 0, 1, 0, 20, 7, 0, 19, 0, 19, 0, 0, 0]\) \\
8 & \( C_0 \cdot (C_2 \cdot x_1 + C_3 + x_1^2)^4 + C_1 \) & \([15, 0, 15, 0, 1, 0, 13, 15, 0, 19, 0, 19, 0, 0, 0]\) \\
8 & \( C_0 \cdot \sqrt{\left|C_2 \cdot \exp(x_0) \cdot \tan(x_0) + C_3\right|} + C_1 \) & \([16, 0, 13, 0, 3, 0, 7, 11, 0, 18, 0, 18, 0, 0, 0]\) \\
8 & \( C_0 \cdot \tan(x_0^2 - \sin(x_0))^2 + C_1 \) & \([15, 0, 7, 0, 2, 0, 5, 15, 0, 18, 0, 18, 0, 0, 0]\) \\
8 & \( C_0 \cdot \left|\left(\frac{x_0^2}{C_3 \cdot x_1 + C_4}\right)^{C_2}\right| + C_1 \) & \([20, 0, 14, 0, 4, 0, 15, 13, 0, 18, 0, 19, 0, 0, 0]\) \\
8 & \( C_0 \cdot \left(C_2 \cdot (\exp(x_0) + \tan(x_1)) + C_3\right)^{\frac{1}{3}} + C_1 \) & \([17, 0, 13, 0, 1, 0, 11, 7, 0, 18, 0, 19, 0, 0, 0]\) \\
8 & \( C_0 \cdot \tan(\tanh(x_1^2 \cdot \sqrt{\left|x_0\right|})) + C_1 \) & \([7, 0, 9, 0, 3, 0, 16, 15, 0, 18, 0, 19, 0, 0, 0]\) \\
8 & \( C_0 \cdot \tanh(\sqrt{\left|C_2 \cdot x_0 + C_3 - \sin(x_0)\right|}) + C_1 \) & \([9, 0, 16, 0, 2, 0, 5, 13, 0, 18, 0, 18, 0, 0, 0]\) \\
8 & \( C_0 \cdot C_2^2 \cdot \exp(\max(0, x_1)) / x_0^2 + C_1 \) & \([3, 0, 15, 11, 0, 6, 0, 21, 0, 18, 0, 19, 0, 0, 0]\) \\
8 & \( C_0 \cdot \tanh(\max(0, x_1)) / \left|x_1\right| + C_1 \) & \([4, 0, 9, 15, 0, 21, 0, 16, 0, 19, 0, 19, 0, 0, 0]\) \\
8 & \( C_0 \cdot (\exp(x_1^2) + \left|x_0\right|^2) + C_1 \) & \([1, 0, 15, 11, 0, 20, 0, 15, 0, 18, 0, 19, 0, 0, 0]\) \\
8 & \( C_0 \cdot \sqrt{\left|\tan(x_1^{C_2} - \tanh(x_1))\right|} + C_1 \) & \([16, 0, 7, 0, 2, 0, 9, 14, 0, 19, 0, 19, 0, 0, 0]\) \\
8 & \( C_0 \cdot \exp(C_2 \cdot x_0^2 \cdot \left|x_0\right| + C_3) + C_1 \) & \([11, 0, 13, 0, 3, 0, 15, 20, 0, 18, 0, 18, 0, 0, 0]\) \\
8 & \( C_0 \cdot \frac{C_2}{\tanh(C_3 \cdot x_1 + C_4 + \tanh(x_0))} + C_1 \) & \([6, 0, 9, 0, 1, 0, 9, 13, 0, 18, 0, 19, 0, 0, 0]\) \\
8 & \( C_0 \cdot \left|\frac{\sqrt{\left|C_2\right|}}{(C_3 \cdot x_1 + C_4)}\right| / \left|x_0\right|^{\frac{1}{4}} + C_1 \) & \([16, 0, 6, 0, 3, 0, 13, 16, 0, 19, 0, 18, 0, 0, 0]\) \\
8 & \( C_0 \cdot \sin\left(\frac{C_2}{-\exp(x_0) + \log\left(\left|x_0\right|\right)}\right) + C_1 \) & \([5, 0, 6, 0, 2, 0, 12, 11, 0, 18, 0, 18, 0, 0, 0]\) \\
8 & \( C_0 \cdot \sqrt{\left|\frac{C_2}{C_3 \cdot x_1 + C_4}\right|} / \left|x_0\right|^{\frac{1}{4}} + C_1 \) & \([16, 0, 6, 0, 3, 0, 13, 16, 0, 19, 0, 18, 0, 0, 0]\) \\
8 & \( C_0 \cdot \tanh\left(\frac{C_2 \cdot \exp(-x_1)}{\sqrt{\left|x_0\right|}}\right) + C_1 \) & \([9, 0, 6, 0, 3, 0, 16, 11, 0, 18, 0, 19, 0, 0, 0]\) \\
8 & \( C_0 \cdot \tanh\left(\frac{C_2 \cdot C_4}{x_0 \cdot \sin(x_1)} + C_3\right) + C_1 \) & \([9, 0, 13, 0, 4, 0, 6, 5, 0, 18, 0, 19, 0, 0, 0]\) \\
8 & \( C_0 \cdot \left|\frac{\tan(x_1)}{x_0^2}\right| + C_1 \) & \([15, 0, 16, 0, 4, 0, 7, 15, 0, 19, 0, 18, 0, 0, 0]\) \\
8 & \( C_0 \cdot \exp\left(\frac{2 \cdot C_2}{x_0} + 2 \cdot \tanh(x_0)\right) + C_1 \) & \([15, 0, 11, 0, 1, 0, 6, 9, 0, 18, 0, 18, 0, 0, 0]\) \\\hline\\[-1.25em]
\label{table: Display Selected Skeleton Information}
\end{longtable}

\end{document}